\documentclass[english,12pt]{elsarticle}
\usepackage[T1]{fontenc}
\usepackage{graphicx}
\usepackage{epstopdf}
\usepackage{epsfig}
\usepackage{subfigure}          
\usepackage{babel}
\usepackage{setspace}
\usepackage{amsmath}
\usepackage{amsfonts}
\usepackage{amssymb}
\usepackage{array}
\usepackage{longtable}
\usepackage{anysize}            
\usepackage{natbib}
\usepackage{algorithm2e}
\usepackage{multirow}
\usepackage[normalem]{ulem}
\usepackage{hyperref}
\usepackage{soul}
\usepackage{lscape}

\makeatletter
\marginsize{3cm}{3cm}{3cm}{3cm}

\begin{document}

\title{Sample Size for Pilot Studies and Precision Driven Experiments}

\author[add1,add2]{C.O.S. Sorzano\corref{cor1}}
\ead{coss@cnb.csic.es}
\cortext[cor1]{Corresponding author}
\author[add1]{D. Tabas-Madrid}
\author[add3]{F. N\'u\~nez}
\author[add4]{C. Fern\'andez-Criado}
\author[add1]{A. Naranjo}

\address[add1]{Centro Nac. Biotecnolog\'ia (CSIC), c/Darwin, 3, 28049 Cantoblanco, Madrid, Spain}
\address[add2]{Univ. San Pablo - CEU, Campus Urb. Montepr\'incipe, 28668 Boadilla del Monte, Madrid, Spain}
\address[add3]{Centro Biolog\'ia Molecular (CSIC), c/Nicol\'as Cabrera, 1, 28049 Cantoblanco, Madrid, Spain}
\address[add4]{Univ. Aut\'onoma de Madrid, Arzobispo Morcillo, 4, 28029, Madrid, Spain}

\journal{Laboratory animals}

\begin{abstract}
   Pilot studies are highly recommended in experiments with animals when little is known about the anticipated values of the mean of the variable under study, its variance or the probability of response. They are also recommended to test the feasibility of the animal model or the experimental technique. However, the sample size required for a pilot study has received little attention and typical sizes practically used go from 5 to 20 animals disregarding any statistical consideration. Pilot studies are a particular case of precision driven experiments in which the sample size is designed according to a desired precision. In this article we provide some statistical guidance on the selection of the sample size of a pilot study whose driving force is the accuracy desired for the determination of the unknown parameters. We provide formulas and design tables for the sample size when trying to determine the standard deviation of a population, its mean, the probability of a certain feature or event, the correlation between two variables, and the survival time before an event occurs. All the calculations performed in this article can also be freely performed on the web through the online calculator available at \url{http://i2pc.es/coss/Programs/SampleSizeCalculator/index.html}. An immediate consequence of our analysis is that typical pilot sizes (5-20) are normally result in a very small precision and that the researcher should be aware of this fact before carrying out a pilot study.
\end{abstract}

\maketitle

\section{Introduction}

Pilot studies constitute the accepted way to acquire basic knowledge about an experimental condition or response that is studied for the first time and about which nothing is known. They are also used to evaluate the viability of an animal model to study a particular experimental condition or to determine the feasibility of some new experimental procedure with animals. When the pilot study is performed with statistical purposes, the goal of a pilot study is normally to provide a range of values for the parameter under study (normally means, standard deviations, proportions, correlations or mean time before an event occurs). This goal is shared by those experimental studies whose main purpose is to provide a confidence interval on a parameter of interest. Although from the statistical point of view, this problem is well known, experimental researchers are not so familiar with it. They are normally used to the standard scientific experimentation in which an hypothesis test is performed at the end of the experiment to ascertain whether the response to the treatment is significantly different from the basal response, or if two bacterial strains significantly differ in their infectious capacities, or if the proportion of respondents has significantly increased. Regulation of animal experimentation has, in the recent years, emphasized the need of a correct statistical design for the approval of the study by the ethical committees. And researchers are slowly incorporating a statistical experimental design (including the sample size) in their applications. However, this design is largely driven by the hypothesis test that ultimately will decide upon the success or not of the research hypothesis. This approach is correct when the experimental setting is relatively known and small variations of previous experiments are addressed. For instance, the response and characteristics of the control group are normally well-known and the desired response of the treatment (the \textit{effect size} in statistical terms) can be easily established with respect to what is known about the control condition and the response to other treatments, strains, etc.

When very little is known about the typical response or size of the variable of interest, researchers feel a little bit lost. They know they need a pilot study, but its size is widely unknown. They simply ``want to see'' the result and they arbitrarily select a number of animals that ranges from 5 to 30 (depending on the experiment cost, its complexity, duration, etc.). For example, a sample size of 12 per group has been suggested as a rule of thumb \citep{Julious2005}. While this number may be appropriate for many experiments (it was derived for the comparison of the means of two groups), it cannot be blindly adopted for any kind of experiment. Normally, researchers are unaware of the statistical consequences of their choice. For instance, if they perform a pilot study with 5 animals to study the proportion of animals responding to a given treatment and 1 of them actually responds (that is an observed proportion of 20\%), then the 95\% confidence interval of the true responding probability goes from 0.5\% to 71.6\% (see Supplementary Table \ref{tab:proportionAccuracy}). That is, with probability 0.95, the true proportion is within this range, which, as can be seen, is not very enlightening about the effectiveness of the treatment (all we know is that it is very unlikely that the large majority, say 90\%, of the population will benefit from it). However, it should be noted that pilot studies are not meant to draw research conclusions. Rather, they are meant to establish approximate ranges of the parameters under study. In some papers \citep{Browne1995b, Sim2012}, it has been suggested to reduce the confidence level from 95\% to a lower value, resulting in a lower number of samples.

With these limitations in mind, in this article we collect the most common cases encountered when designing a pilot study. By far, the three most typical objectives, from the statistical point of view, to design a pilot study are to give an acceptable range for the standard deviation, or mean of a variable of interest, or for the proportion of individuals exhibiting a given characteristic. Less common are the pilot studies whose objective is to determine the correlation of a given variable, or the mean time before an event (infection, cure, death, ...) occurs (for space reasons, these latter two cases have been moved to the Supplementary Material). However, for completeness, we have also included these two cases in the article. The aim of the article is double. On one hand, it aims at serving as a one-stop reference for the design of pilot studies (many other articles have addressed specific aspects of this problem, but the information is scattered in different articles and even scientific domains \citep{Kieser2000, Lenth2001, VanTeijlingen2002, Lancaster2004, Naing2006, Hertzog2008, Arain2010, Viechtbauer2015, Whitehead2016}). Given the desired accuracy and confidence level, we provide tables with the required number of samples. In most cases the accuracy is specified as a percentage of the magnitude to be estimated and not in natural units. We realize that many of these sample sizes are unrealistically large for a pilot study. However, this is intentionally done to make the researcher aware of the constraints imposed by a too ambitious design. On the other hand, the second aim of the article is to give feedback to the researcher on the precision obtained if typical pilot sizes (between 5 and 20) are chosen. There is nothing wrong with these sizes as long as we are aware of their accuracy and this accuracy suffices for our research purposes.

It should also be noted that the sample sizes calculated in this article are based on precision requirements. In this way, the sample size is independent of the animal species under experimentation (an experiment with 8 mice is typical, with 8 zebra fish would be considered clearly insufficient, and with 8 primates would normally be considered a large experiment). However, from the statistical precision point of view, they all have the same precision (note that in our approach the precision acts normally as a multiplier of the observed variability; meaning that any experimental procedure aiming at reducing this variability helps to reduce the confidence interval and our uncertainty about the underlying true statistical parameter). A different issue is the similarity of the species to human or the relevance of results in that animal model. Actually, to increase the relevance of the experimental findings it is normally recommended to use different strains of the animal model rather than increasing the sample size using only one strain.

We provide formulas for the calculation of the sample size when the experiment objective is determining a confidence interval for the standard deviation, the mean, the proportion, the correlation between two variables and the mean lifetime. Some of these formulas are not easy to calculate by hand (although some approximations are relatively simple to handle and good enough to have an order of magnitude, that is a value in the same range as the exact value computed from the more complicated formula). We provide tables using those formulas trying to cover the most common desired accuracies and confidence levels. We have calculated these tables with the help of the program Pass 14 (PASS 14 Power Analysis and Sample Size Software (2015). NCSS, LLC. Kaysville, Utah, USA, ncss.com/software/pass) and they can be freely replicated at the online calculator at \url{http://i2pc.es/coss/Programs/SampleSizeCalculator/index.html}, which has specifically been designed for this article. 

For each experimental case, we provide three sections: 1) Typical scientific questions, in which we pose some questions that try to help the researcher to identify his/her case; 2) Statistical derivation (in the Supplementary Material), in which we give the formulas required for the sample size calculation (practitioners may skip this part, although it is recommended to know the assumptions under which the sample size has been calculated and under which it gives a good answer to the experimental problem); 3) Worked example, in which we use the tables for each case in order to illustrate an example of how to use them. For each experimental case in the Supplementary Material we provide all the details about the assumptions performed during the derivation of the sample size formula. One of the key assumptions in many of them is the normality of the observed data. Small violations of this assumption does not rend useless the calculated sample size. These violations cause a distortion in the confidence level associated to the sample size (for instance it may shift from the typical 95\% to 90\% or 97\%, depending on whether the true underlying population has fatter or narrower tails than the Gaussian). Large violations of the normality assumption may severely compromise the sample size calculation.

\section{Animals}

Not applicable.

\section{Materials and Methods}

\subsection{Pilot study to determine a standard deviation}

\underline{Typical scientific questions:}
\begin{itemize}
	\item What is the variability of a gene expression level under a specific condition?
	\item What is the variability of mice or strain response to a certain drug?
\end{itemize}

Many pilot studies aim at determining a sensible range of the variance of a given variable when there is no prior information about it. Variance is the square of the standard deviation, so talking about variance and standard deviation is about the same concept; however, computing confidence intervals for the standard deviation has the advantage that the standard deviation has the same units as the measurements (and not squared units), and makes the experimenter more familiar with the absolute values she is used to observe. How many samples do we need to determine its standard deviation with 10\% of accuracy and a level of confidence of 95\%? We will refer to the 10\% of accuracy as $\delta$ (=0.1) and the confidence level will be referred to as $1-\alpha$ (for instance, if the confidence level is 95\%, then $\alpha=0.05$). As shown in the Supplementary Material, the relationship between the sample size, $N$, and the precision is
\[\delta=\sqrt{\frac{N-1}{\chi^2_{\frac{\alpha}{2},N-1}}}-1\]
and the confidence interval for the true standard deviation, $\sigma_0$, is
\[\begin{array}{c}
	 \sigma_0 \in \left[s\sqrt{\frac{N-1}{\chi^2_{1- \frac{\alpha}{2},N-1}}}, s\sqrt{\frac{N-1}{\chi^2_{\frac{\alpha}{2},N-1}}}\right]
\end{array}\]
where $\chi^2_{p,N-1}$ is the $p$-th percentile of the $\chi^2$ distribution with $N-1$ degrees of freedom.

	Supplementary Table \ref{tab:varianceSampleSize} shows the sample size for some common values of $\delta$ and $\alpha$.
	Conversely, we may calculate the accuracy for some typical pilot sample sizes (Supplementary Table \ref{tab:varianceAccuracy}). 

\underline{Worked example:}

In our example, to achieve such a high accuracy ($\delta=0.1$) we would need 234 samples (Supplementary Table \ref{tab:varianceSampleSize}). This number is way too far from the commonly accepted pilot studies, revealing how unrealistic was our original aim of determining the standard deviation with a precision of 10\%. If we are willing to use only $N=5$ animals in our pilot study,
	then the accuracy would drop to $\delta=187.36\%$ (Supplementary Table \ref{tab:varianceAccuracy}). This means that the confidence interval $[0.5991s,2.8736s]$ contains the true standard deviation with probability 95\%. That is, one of the confidence limits can be as far as $1+\delta=2.8736$ away from the sample standard deviation. This is the meaning of an accuracy larger than 100\%, the accuracy is a number that helps to construct the upper limit, in the case of the standard deviation, of the confidence interval; as such, this number can, without any inconsistency, be larger than 100\%. Note that the confidence interval is not symmetric, meaning that the confidence interval is not $s(1\pm \delta)$, which would result in a negative lower limit that does not make sense for a standard deviation.
	
Remark that this confidence interval is defined using the observed variability. In this way, any experimental technique aimed at reducing the observed variance (blocking variables, more precise measurements, etc.) will help to reduce the uncertainty about the true underlying variability. This result is also independent of the type of animals being studied (outbred, inbred, different strains). In practice, different strains have different variability. This will result in a different observed standard deviation, $s$, which is used to construct the confidence interval. Note that our choice of the sample size, $N$, determines the coefficients that multiply $s$. Larger $N$s imply multiplication coefficients closer to 1, so that the confidence interval is narrower.

\subsection{Pilot study to determine a mean}

\underline{Typical scientific questions:}
\begin{itemize}
	\item What is the mean gene expression level under a specific condition?
	\item What is the mean response of mice or a strain to a certain drug?
\end{itemize}

In this kind of experiments, the goal is not to determine the variability of a random variable, but its mean value. As can be seen in the statistical derivation in the Supplementary Material, the most convenient way of designing these pilot studies is by choosing an accuracy that is a fraction of the sample standard deviation. As shown in the Appendix, the relationship between the sample size, $N$, and the precision, $\delta$, is
\[\delta=\frac{t_{1-\frac{\alpha}{2},N-1}}{\sqrt{N}}\]
The confidence interval for the true mean, $\mu$, is 
\[\mu \in \left[\hat{\mu}-s\delta, \hat{\mu}+s\delta\right]\]
where $1-\alpha$ is the confidence level of the confidence interval, $\hat{\mu}$ the sample mean and $s$ the sample standard deviation. Note that $\delta$ represents a fraction of the sample standard deviation. For instance, $\delta=0.2$ implies that our goal is to determine the mean with a precision that is 20\% of the standard deviation of the population being studied. In the case of the mean, the confidence interval is symmetric ($\hat{\mu}\pm s\delta$). However, this is not the general case for most statistical parameters.

Supplementary Table \ref{tab:meanSampleSize} shows the sample size for some common values of $\delta$ and $\alpha$. Conversely, we may calculate the accuracy for some typical pilot sample sizes (Supplementary Table \ref{tab:meanAccuracy}). 

\underline{Worked example:}

As an example, let us assume we want to determine the population mean with a precision of $\delta=0.2$ and a confidence level of $95\%$, then we need $N=99$ animals in our experiment. Alternatively, if we use $N=5$ animals, then the maximum achievable accuracy with the same confidence level is $\delta=124.17\%$ (this means that, with a probability of 95\%, the true mean is at most 124.17\% times the sample standard deviation away from the sample mean). Note that our precision is defined as a fraction of the sample standard deviation. In this way, any experimental action aimed at reducing the sample variability (controlling the animal strain or the experimental conditions) will result in narrower confidence intervals.
	
\subsection{Pilot study to determine a paired difference}

\underline{Typical scientific questions:}
\begin{itemize}
	\item What is the mean difference between the blood pressure of an animal with hypertension before and after a new treatment?
	\item What is the mean difference between the weight of several pairs of homozygous mice when they are given two different diets?
\end{itemize}

In these experiments, an animal (or a genetically identical animal in the same condition because environment has an influence on the response) serves as its own control and the goal is to determine the effect size of a given treatment. Sometimes, paired data is acquired using the same strain with and without treatment. However, to effectively serve as control, all experimental conditions between the two groups of animals should be as similar as possible (same diet, same technical staff, both experiments performed in the same days, ...). Reference values for that strain do not serve as control because this data values have been obtained in a different laboratory with many differing variables (diet, staff, circadian rythms, ...). These experiments can be designed following the same principles as the experiments trying to determine a mean. The key is to note that the variable of interest is the difference between two random variables:
\[D=X_1-X_2\]
where $X_1$ is the blood pressure before the treatment and $X_2$ after the treatment, for example. We can calculate this difference for each animal (analogusly, each homozygous pair in the example with two diets) participating in the experiment. The variable $D$ has a true mean difference, $\mu_D$, and an observed mean difference, $\hat{\mu}_D$. Let us call as $s_D$ to the observed standard deviation. The accuracy, $\delta$, is defined exactly in the same way as in the previous section (Pilot study to determine a mean), and the confidence interval is also similar
\[\mu \in \left[\hat{\mu}_D-s_D\delta, \hat{\mu}_D+s_D\delta\right]\]
This time, $N$ is the number of pairs of measurements we need to collect. If the two measurements are obtained from the same animal, then $N$ is the total number of animals. If the two measurements are obtained from two genetically identical mice, then $N$ is the number of paired groups. 

Supplementary Tables \ref{tab:meanSampleSize} and \ref{tab:meanAccuracy} are also valid for the case of paired samples. 

\underline{Worked example:}

As an example, let us assume we want to determine the effect of a drug on blood pressure. For doing so, we will measure the blood pressure of an animal before and after the drug administration. We want to determine the population mean difference with a precision of $\delta=0.2$ and a confidence level of $95\%$, then we need $N=99$ animals in our experiment (each animal will be measured twice, once before the treatment and once after the treatment). Alternatively, if we perform the pilot study with $N=5$ animals, then the maximum achievable accuracy with the same confidence level is $\delta=124.17\%$ (this means that, with a probability of 95\%, the true mean difference is at most 124.17\% times the sample standard deviation away from the sample mean difference).

\subsection{Pilot study to determine a proportion expected to be between 10\% and 90\%}

\underline{Typical scientific questions:}
\begin{itemize}
	\item We want to determine, with a confidence level of 95\%, which is the proportion of a population of mice having a particular genetic variant (we expect this proportion to be about $p=15\%$) with a precision of $\delta=5\%$ (i.e., the true proportion is within a range of $\pm 5\%$ about the observed proportion with probability 95\%).
	\item We want to determine, with a confidence level of 95\%, which is the proportion of a population of mice with a particular kind of cancer to survive after 6 months.
\end{itemize}

Lethal dose 50 (the dose at which 50\% of the animals die by toxicity) does not fall within this experimental design. Normally these studies are performed by a sequential increase/decrease of the dose and the statistical design in this case follows a different reasoning \cite{Lipnick1995,Rispin2002}.

As shown in the Supplementary Material, the exact confidence interval for the true proportion, $p$, when the observed proportion is $\hat{p}$ is
\[p\in\left[
\frac{\hat{p}N}{\hat{p}N+(N-\hat{p}N+1)F_{1-\frac{\alpha}{2};2(N-\hat{p}N+1),2\hat{p}N}},
\frac{(\hat{p}N+1)F_{1-\frac{\alpha}{2};2(\hat{p}N+1),2(N-\hat{p}N)}}{(N-\hat{p}N)+(\hat{p}N+1)F_{1-\frac{\alpha}{2};2(\hat{p}N+1),2(N-\hat{p}N)}}
\right]\]
where $F_{y;m,n}$ is the $y$-th percentile of Snedecor's F with $m$ and $n$ degrees of freedom. The exact expression does not easily lend itself to give a clear relationship between the sample size and the precision. Under certain conditions (see Supplementary Material), this confidence interval can be approximated by the interval
\[p\in\left[\hat{p}-z_{1-\frac{\alpha}{2}}\sqrt{\frac{\hat{p}(1-\hat{p})}{N}},\hat{p}+z_{1-\frac{\alpha}{2}}\sqrt{\frac{\hat{p}(1-\hat{p})}{N}}\right]\]
With this approximation, the precision is 
\[\delta=z_{1-\frac{\alpha}{2}}\sqrt{\frac{\hat{p}(1-\hat{p})}{N}}\]
It may sound counter-intuitive that before performing the pilot study that aims at determining the sought proportion, we need to assume in which range this proportion may be. Unfortunately, unlike the designs for the standard deviation and the mean, the design formulas need to make a prior assumption about the expected value of the observed proportion, $\hat{p}$. If no prior information exists at all about the expected proportion, then we may design the experiment in the worse case with $\hat{p}=0.5$ (note that at this value we have the least precision).

Supplementary Table \ref{tab:proportionSampleSize} shows the sample size for some common values of $\delta$ and $\alpha$. Conversely, we may calculate the accuracy for some typical pilot sample sizes (Supplementary Table \ref{tab:proportionAccuracy}). 

\underline{Worked example:}

In our example, the genetic variant is presumed to be about $p=15\%$ and the desired precision $\delta=5\%$. Assuming a confidence level of 95\%, we look up Supplementary Table \ref{tab:proportionSampleSize} to realize that we need to check 215 animals (if we now perform the experiment with 215 animals and we observe a proportion of $\hat{p}=15\%$, then the 95\% confidence interval will be between 10\% and 20\%). If we cannot afford more than $N=20$ animals, and the observed frequency is $\hat{p}=15\%$, then with probability 95\%, the true probability is within the range [3.2,37.9]\% (Supplementary Table \ref{tab:proportionAccuracy}), that is the expected precision with only 20 animals is about $\delta=37.9-15=22.9\%$.

\subsection{Pilot study to determine a proportion expected to be below 10\% or above 90\%}

\subsubsection*{Two-sided intervals}

\underline{Typical scientific questions:}
\begin{itemize}
	\item What is the prevalence of a rare disease in a population of animals?
\end{itemize}

For rare diseases or infrequent adverse effects, the design tables of the previous section are of little use, because the precision in the determination of the proportion is well beyond the proportion itself (for instance, there is little use in determining the prevalence of a rare disease as $\hat{p}=1\%$ if the precision is $\delta=5\%$, that is, the confidence interval for the prevalence is [0,6]\% while the prevalence estimate itself is 1\%; the same applies for very frequent features, $p=99$\% whose confidence interval would be [94,100]\%; an application example would be animals that are normally resistant to a bacterial strain). In these cases, it is better to construct a confidence interval whose precision is a fraction of the expected frequency, $\delta=kp$ (with $0<k<1$, in our example it is much more useful to say that the prevalence is in the range [0.5,1.5]\%) \citep{Naing2006}. This is an example of a two-sided confidence interval. The exact formula in the section above are still valid (the Gaussian approximation may be not if the sample size is not large enough).

Supplementary Table \ref{tab:smallProportionSampleSize} shows the sample size for some common values of $k$ and $\alpha$. Conversely, we may calculate the accuracy for some typical pilot sample sizes (Supplementary Table \ref{tab:smallProportionAccuracy}). 

\underline{Worked example:}

In the example above, in which we wanted to determine the prevalence of a rare disease (expected to be about 1\%) with a precision of 0.5\% and a confidence level of 95\%, we need to check 1,741 animals (see Supplementary Table \ref{tab:smallProportionSampleSize} and note that 0.5\% is 50\% of the nominal expected proportion 1\%). This huge number is a direct consequence of a very low prevalence.

\subsubsection*{One-sided intervals}

\underline{Typical scientific questions}:
\begin{enumerate}
	\item We are developing a new vaccine and we want to know an upper bound of the proportion of animals that get infected when they are exposed to the pathogen. We expect this upper bound to be smaller than 1\%.
	\item We want to know an upper bound for the proportion of animals in our animal facility with a given bacterial strain in their guts. We expect this upper bound to be smaller than 2\%.
	\item Can I detect an infection in my animal facility (regular checks are needed for disease controls)? It is assumed that all cages have the same infection probability.
	\item Our animal experiment must be carried out by a third party. We want to estimate an upper bound of the proportion of samples that will be spoiled by a misunderstanding of the instructions. We expect this proportion to be below 5\%.
	\item We want to know an upper bound for the proportion of animals having adverse effects with a given drug. We expect this upper bound to be smaller than 1 animal in 1000.
	\item We want to know a lower bound for the proportion of animals that are correctly labeled in our animal facility. We expect this lower bound to be larger than 99\%.
\end{enumerate}

In these examples, we are interested in one-sided confidence intervals. The construction of these intervals requires a different mathematical treatment and, as in the two previous sections, we also need a prior assumption about its expected value. In the Supplementary Material we provide formulas for low proportions (as in the example of the vaccine, the proportion of infected animals) and high proportions (as in the example of correctly labeled animals). However, we can always turn a problem with high proportion into one of low proportion (for instance, by studying the proportion of incorrectly labeled animals instead of correctly labeled ones). For simplicity, we will present the low proportion case in the main text and both cases in the Supplementary Material. As shown in the Supplementary Material, if the expected proportion, $p$, is very low the exact confidence interval is
\[p\in \left[0,\frac{(\hat{p}N+1)F_{1-\alpha;2(\hat{p}N+1),2(N-\hat{p}N)}}{(N-\hat{p}N)+(\hat{p}N+1)F_{1-\alpha;2(\hat{p}N+1),2(N-\hat{p}N)}}\right]\]

This formula is inconvenient for manual calculations. If $\hat{p}=0$, that is we have not observed any animal with the studied property (infection, incorrect label, ...), then the confidence interval can be calculated (exactly) as
\[p\in\left[0,1-\alpha^{\frac{1}{N}}\right]\]
If we are designing the experiment, it is easy to derive
\[N\geq \frac{\log(\alpha)}{\log(1-p_U)}\]
where $p_U$ is the expected upper bound of the proportion (that is the confidence interval is expected to be $[0,p_U]$). This strategy is called zero-acceptance experiment and it is the standard procedure for instance for disease controls in animal facilities.
This formula can be easily calculated. However, in Epidemiology there is a widespread approximation known as the \textit{rule of three} based on a $\chi^2$ approximation to the zero-acceptance problem. The result is
\[N\geq\frac{2.9957}{p_U}\]
which is the well-known rule of three. Remind that this rule was derived under the assumption that we will observe $N$ animals, if none of them is infected, then we accept the alternative hypothesis.

Supplementary Table \ref{tab:smallProportionOneSidedSampleSize} shows the sample size required for one-sided intervals and several typical limits. 

\underline{Worked example:}

In the example of the new vaccine, in which we expected the proportion of infected animals to be below 1\%, the 95\% confidence interval constructed after the experiment will have this precision if we use $N=299$ animals and we do not observe any infection in this group. 

For disease controls it is typical to use 10 animals. This would allow us to construct a confidence interval $[0,0.2996]$, that is if we observe 0 infections amongst the 10 animals, the disease may still have a prevalence in the animal house as high as 29.96\% and simply due to bad luck in our sampling we did not observe any of the infected animals. In practice this situation is ameliorated because the cage beds from many cages are used to house for a period the 10 animals to be tested in the disease control. In this way, if there is an infection in the animal facility, there is a higher chance to be detected (in a way, these 10 animals are representative of a larger number of animals). However, this is not an ideal situation because not all infections are equally transmitted through the cage bed.

\section{Results}

Not applicable.

\section{Discussion}

The sample size required for a pilot study has been designed in this article as a function of the desired accuracy and the confidence level with which we want to construct a confidence interval containing the true parameter of the underlying distribution. More accurate and confident intervals will require larger sample sizes. In many cases, the sample size for a desired accuracy and confidence level becomes unrealistically high. And in many of these occasions researchers will most likely opt for a typical pilot size somewhere between 5 and 20 animals. This is statistically correct as long as researchers are aware of the accuracy achieved with their sample size. We have given tables showing the accuracy achieved for different situations (pilot study for the standard deviation, for the mean, proportions, ...). As long as the accuracy achieved suffices for the research purposes, there is no objection in its use. However, there are situations like estimating very low proportions or relatively weak or moderate correlations in which the resulting confidence interval is almost useless (not much is known about the true underlying proportion or correlation after the experiment has been finished).

This article provides a statistical guide for the design of pilot studies, which in many applications to the ethical committees seem to be totally unguided under the assumption that just a few animals (5-20) suffices, in all circumstances, for the purpose of having an estimate of any parameter of interest. We have highlighted that depending on the expected results, this number may be too low resulting in a waste of research budget and animal lives. Actually, the statistical information gained by a small number of animals is in some cases so small that one may wonder whether pilot studies are ethically justified and a similar level of precision could not be achieved by analysis of preexisting experiments reported in the scientific literature. On the other side, pilot studies may be justified to gain familiarity with the experimental procedure and material, to verify the logistics for a future larger study, to check that the staff is sufficiently trained to perform the procedures, that the instructions are clear, and that all the steps proposed are feasible. A different story are those studies, whose aim is to set a confidence interval on a parameter under study, always bearing in mind a tradeoff between the required precision and the ethical cost of carrying out an experiment whose consequences for the animal lives are not negligible.

All the calculations performed in this article can also be freely performed on the web through the online calculator available at \url{http://i2pc.es/coss/Programs/SampleSizeCalculator/index.html}.

\section*{References}


\newpage

\section*{Supplementary Material}

\subsection*{Pilot study to determine a correlation}

\underline{Typical scientific questions}:
\begin{enumerate}
	\item Is there any relationship between the expression level of this gene and blood pressure?
	\item Is there any relationship between air contaminants and breath capacity?
\end{enumerate}

\begin{figure}
	\centering
	\includegraphics[width=5cm]{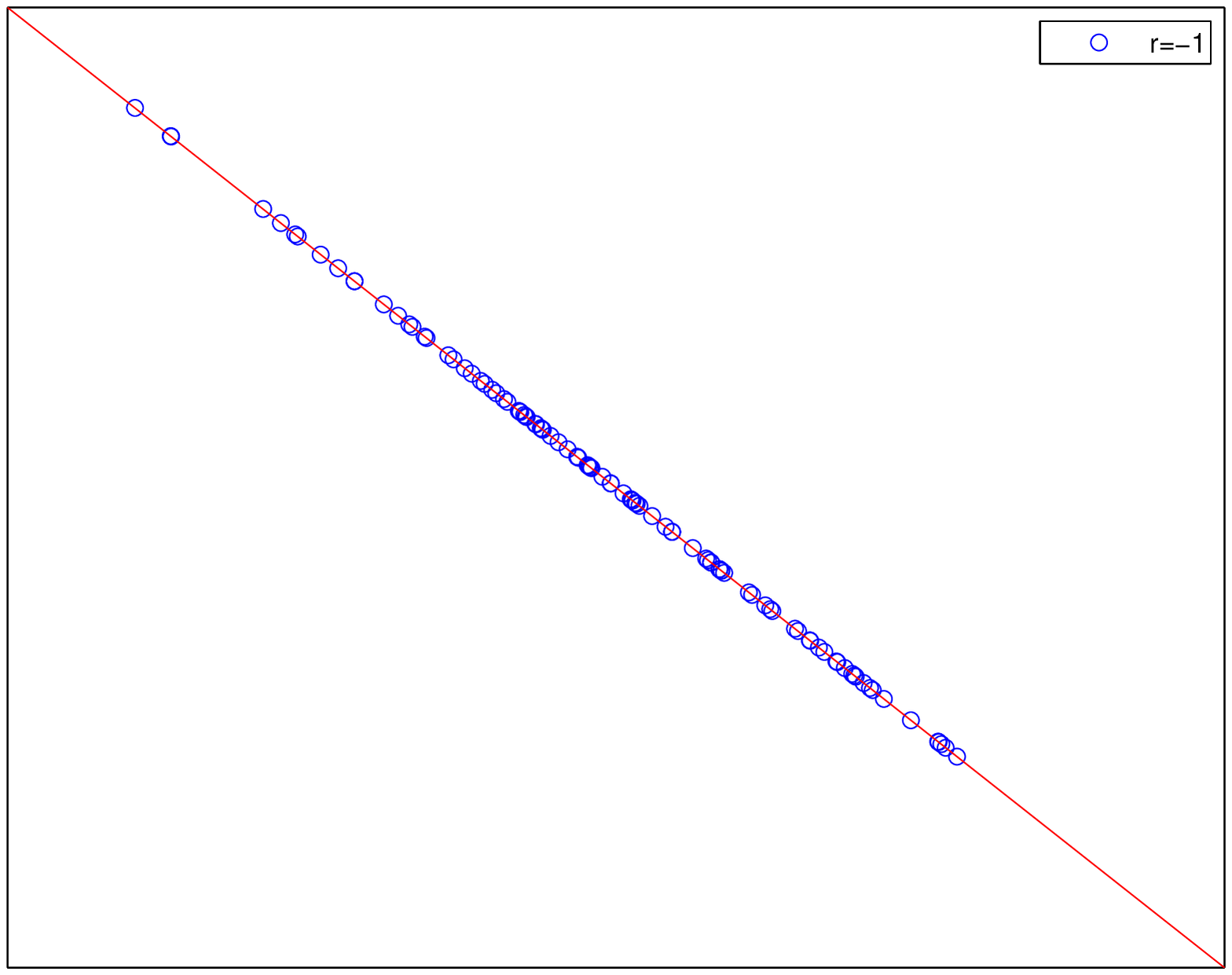}
	\includegraphics[width=5cm]{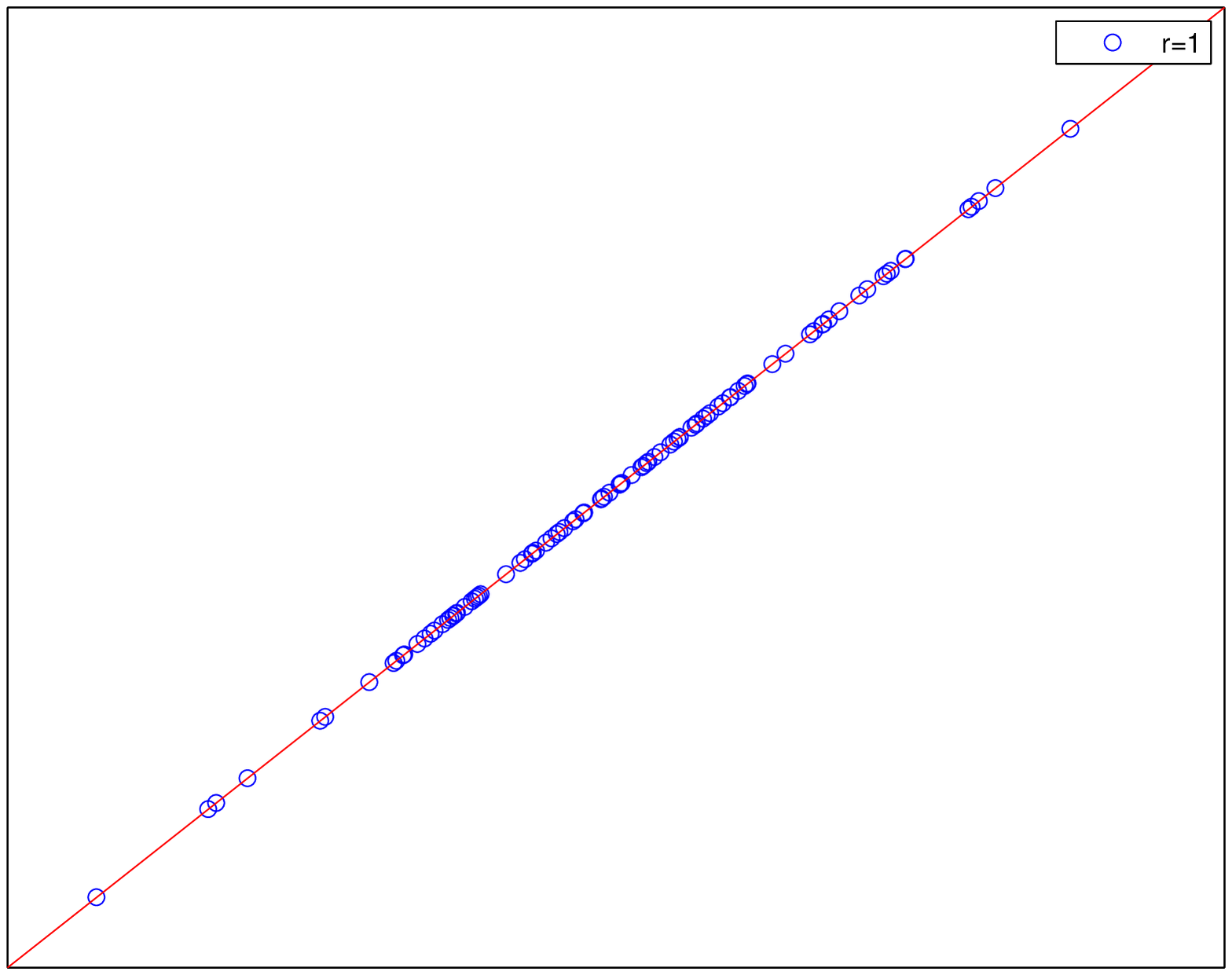} \\
	\includegraphics[width=5cm]{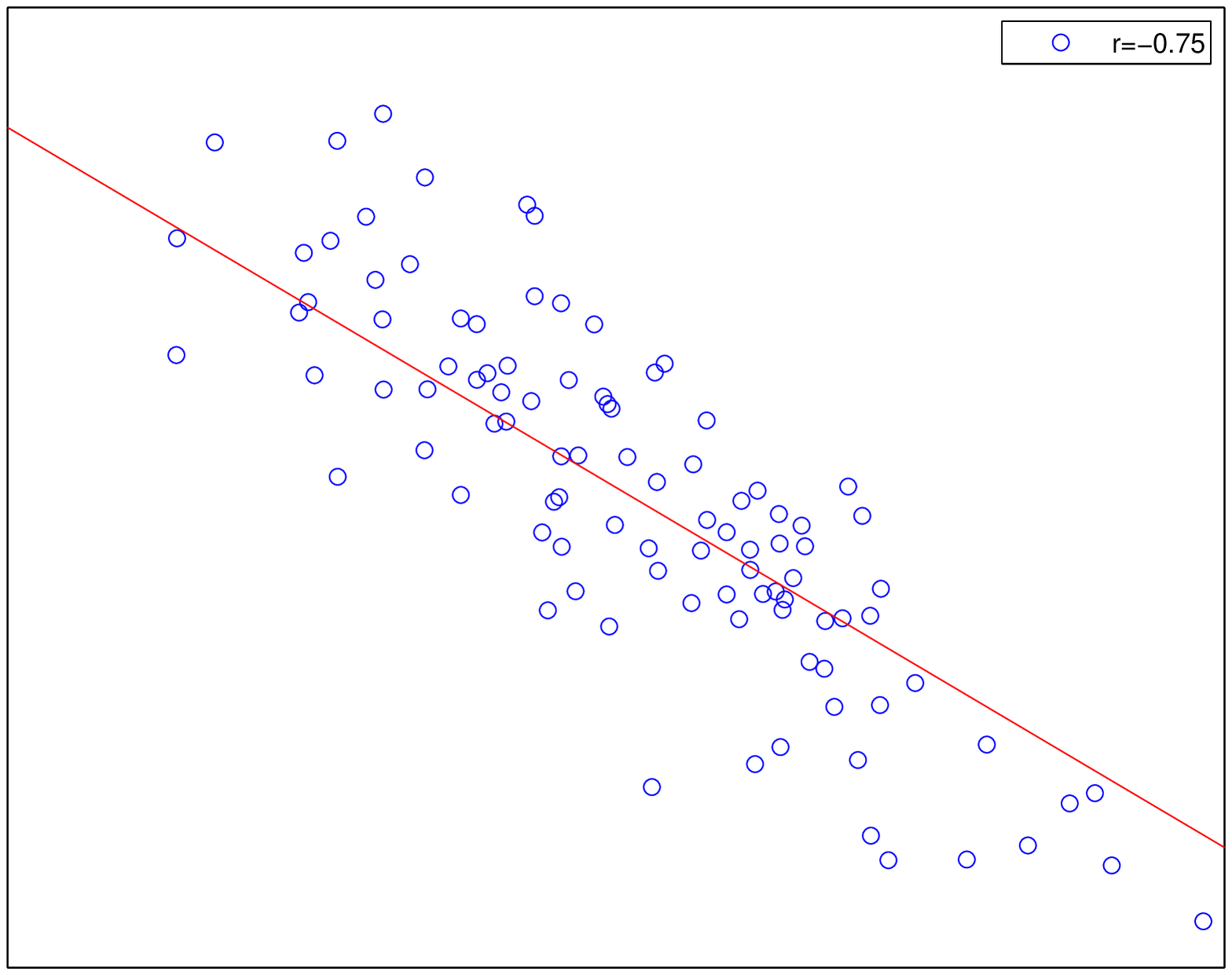}
	\includegraphics[width=5cm]{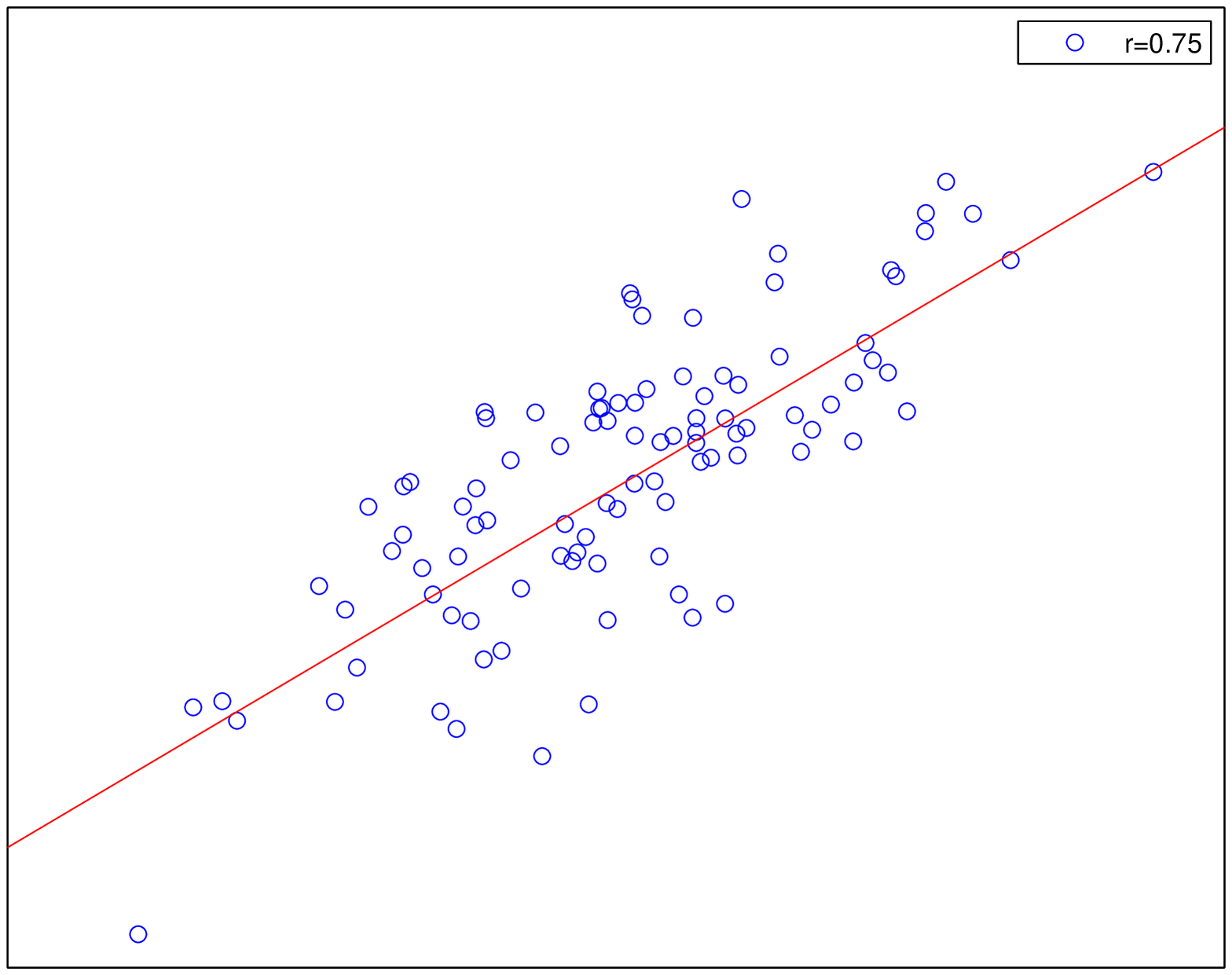} \\
	\includegraphics[width=5cm]{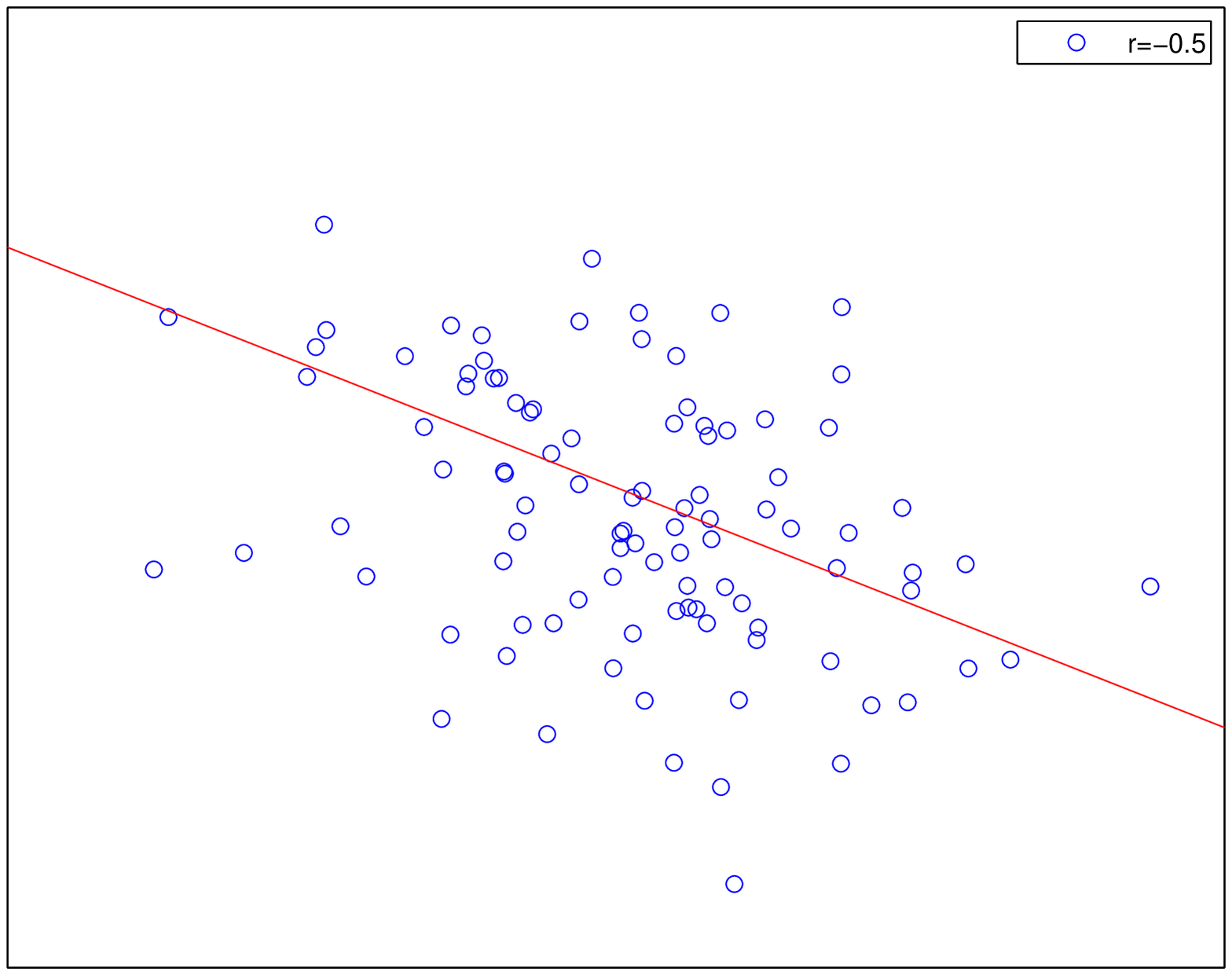}
	\includegraphics[width=5cm]{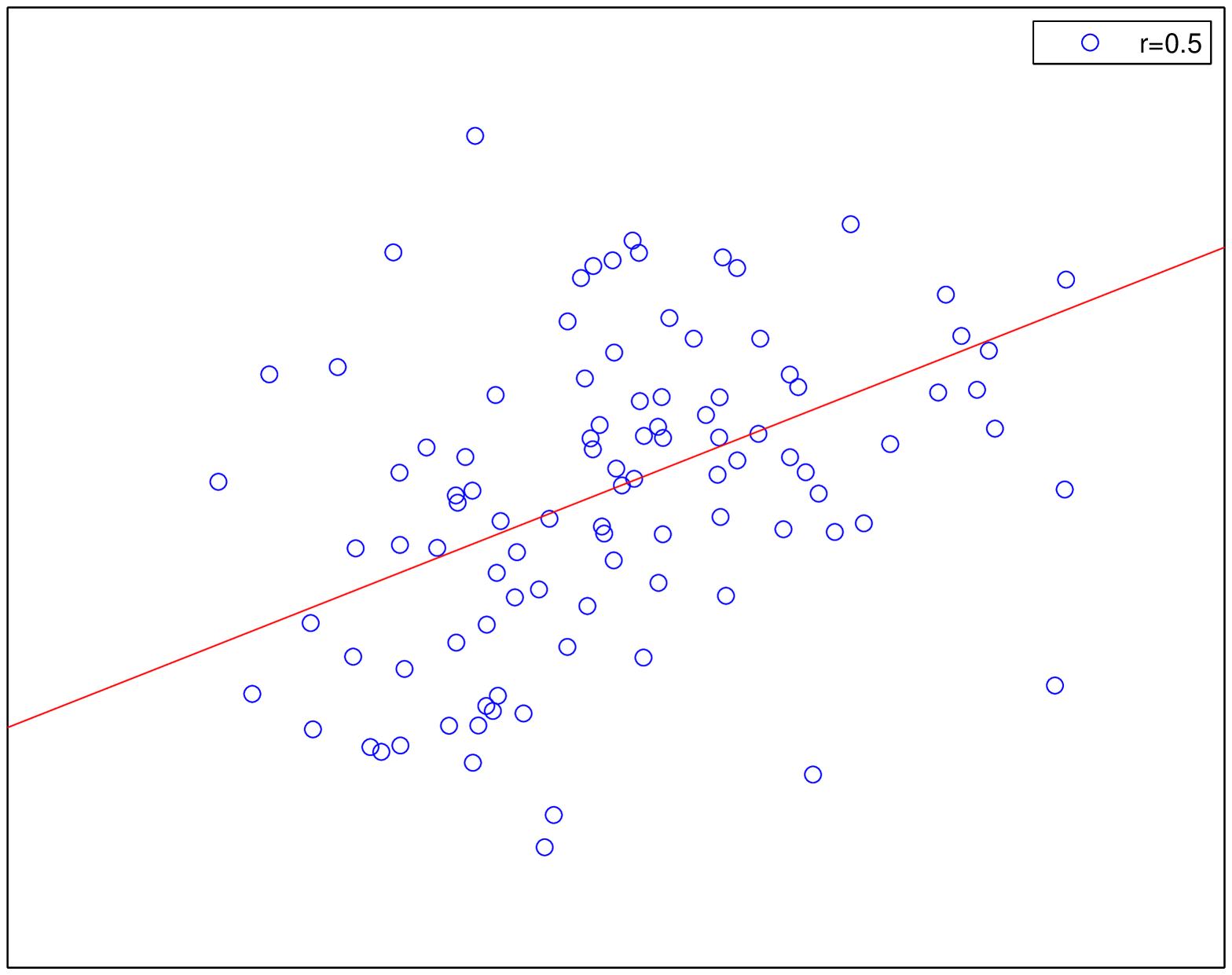} \\
	\includegraphics[width=5cm]{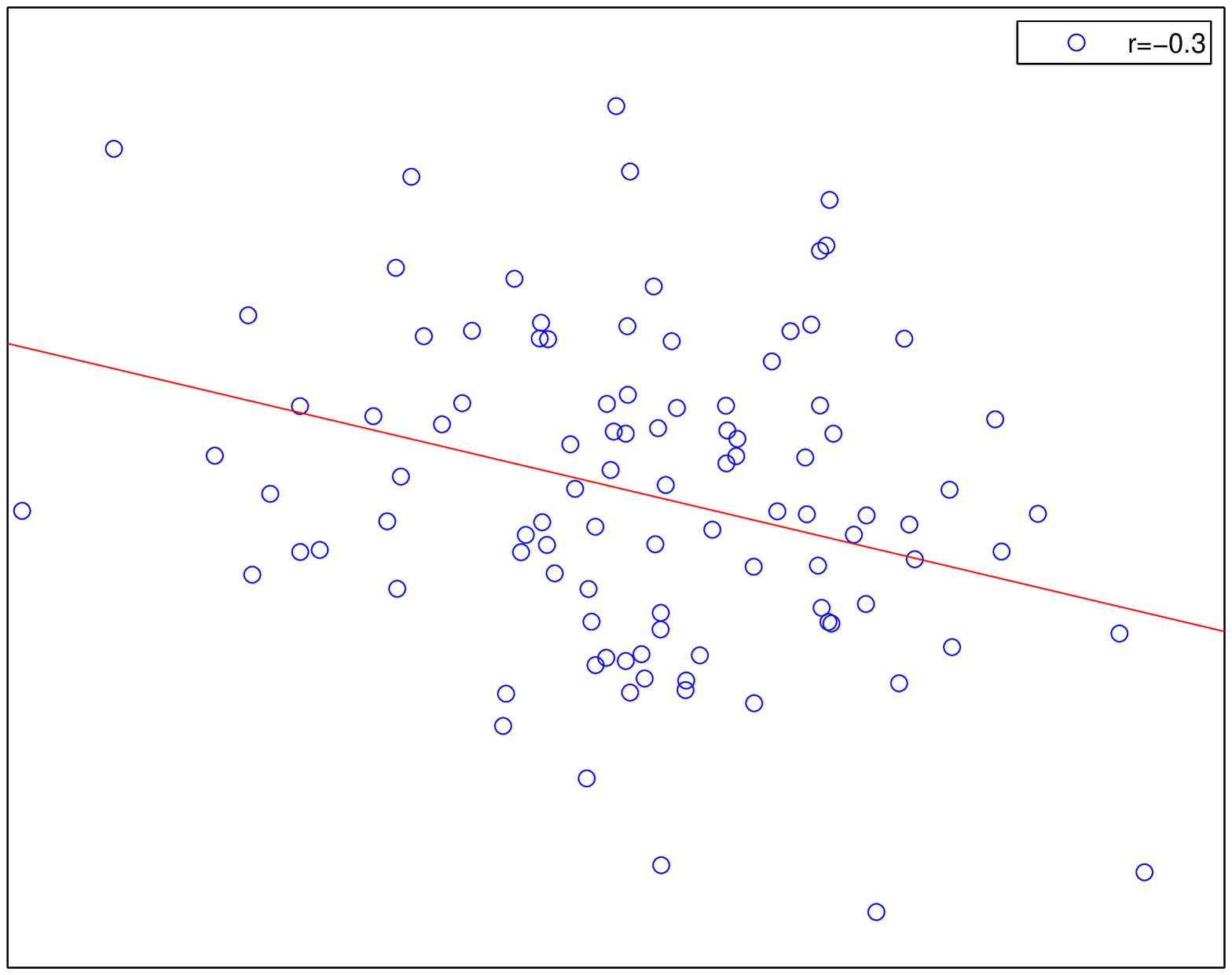}
	\includegraphics[width=5cm]{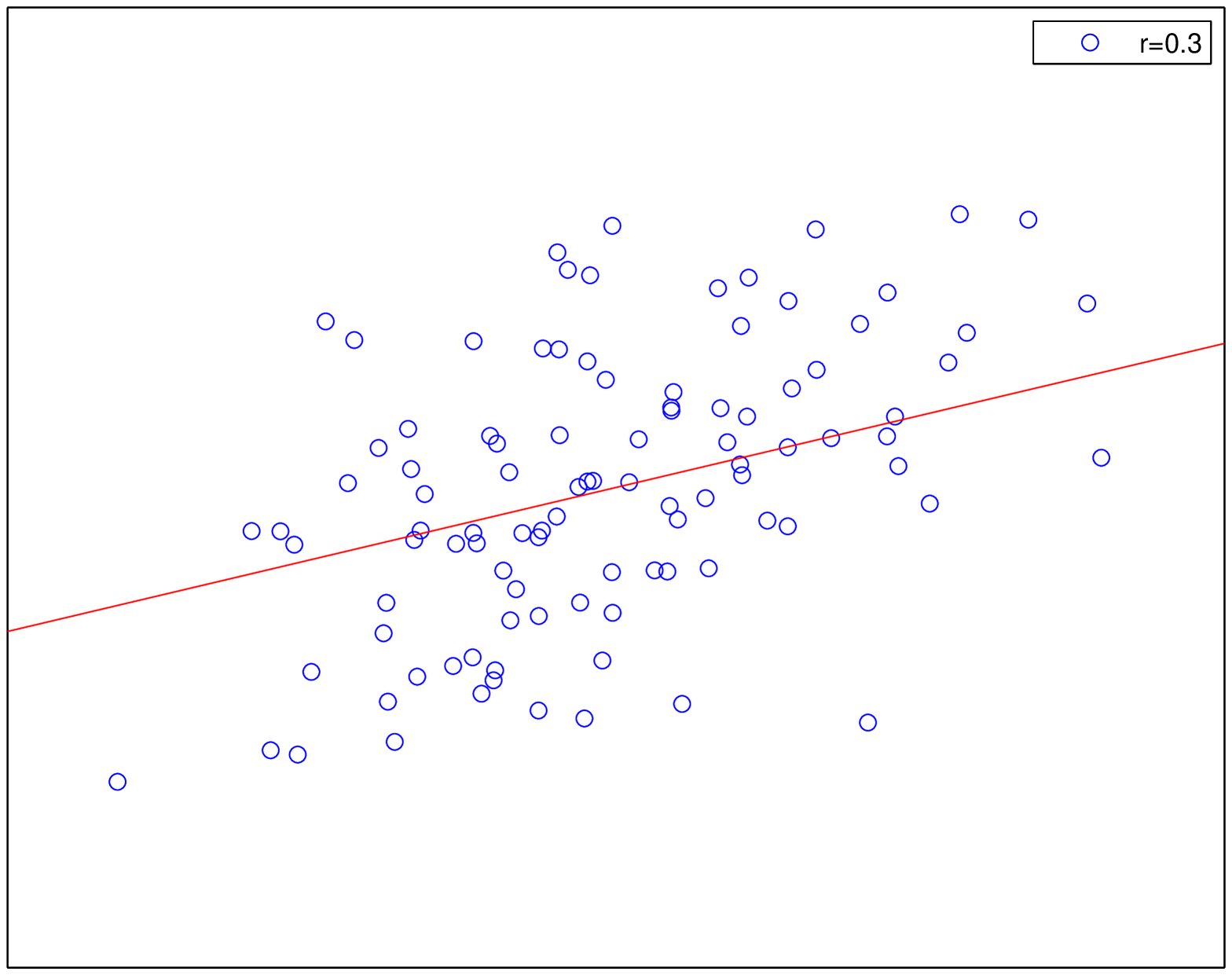} \\
	\includegraphics[width=5cm]{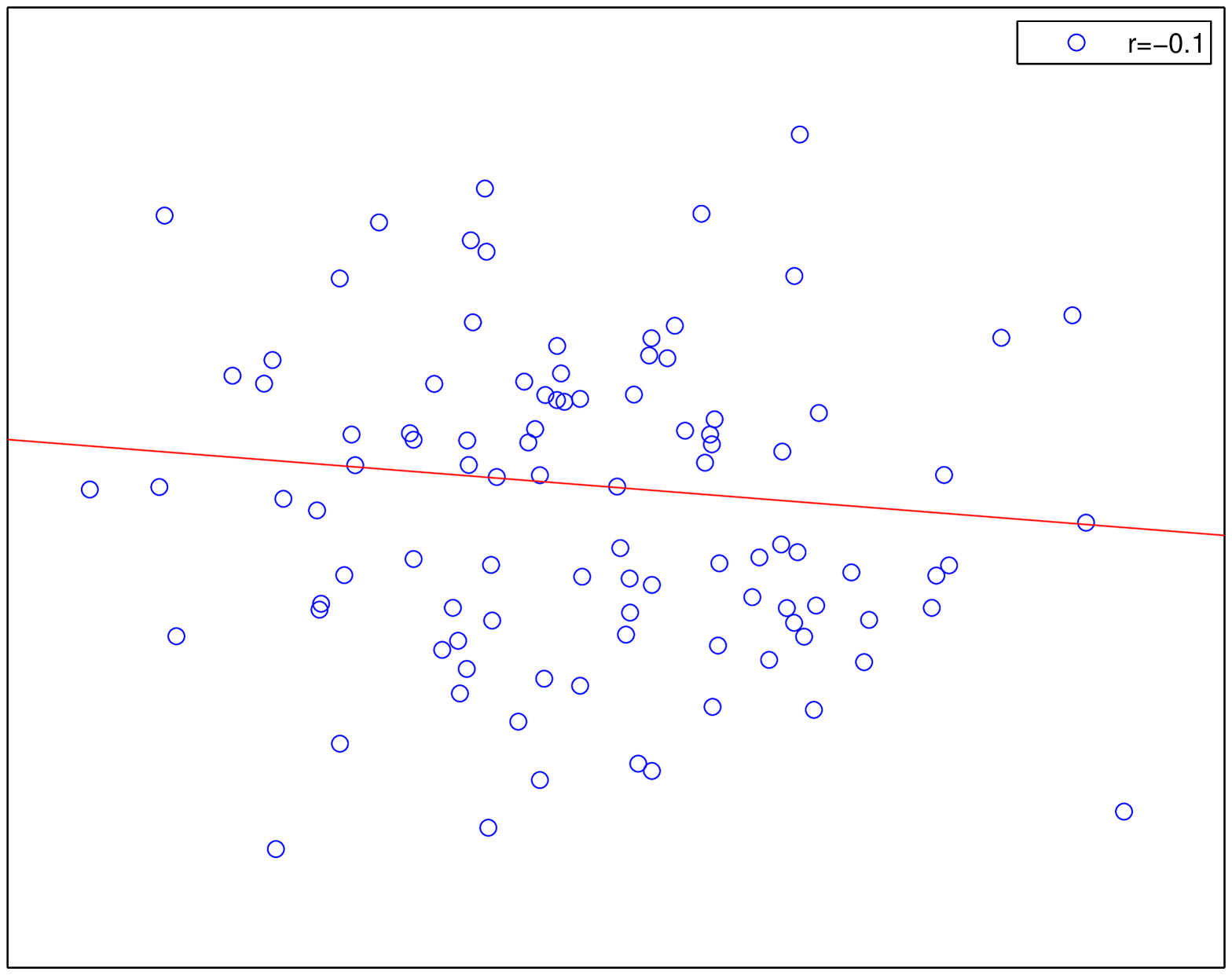}
	\includegraphics[width=5cm]{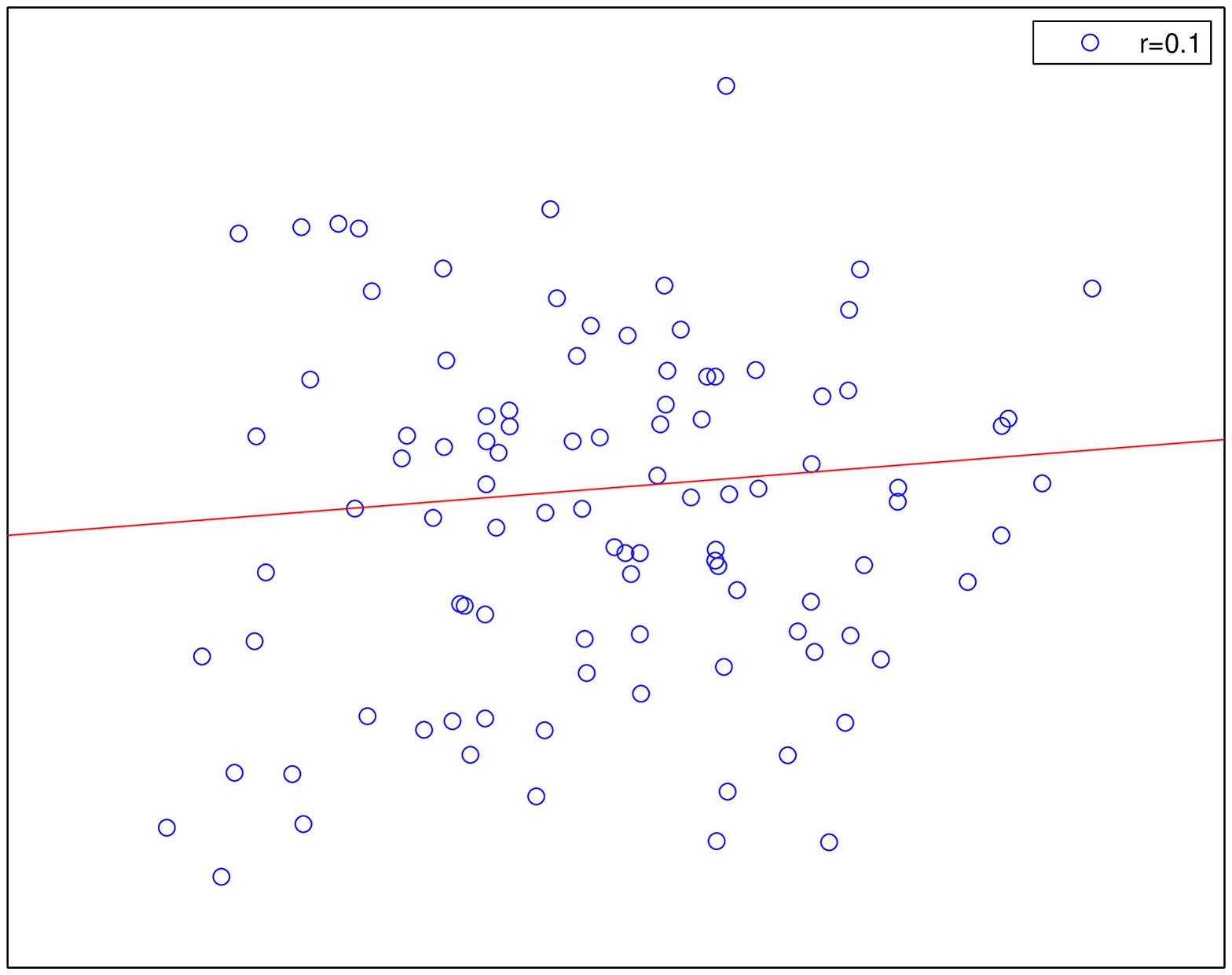} \\
	\caption{Examples of datasets with different correlation coefficients (1, 0.75, 0.5, 0.3, 0.1). Left column: negative correlation coefficients. Right column: positive correlation coefficients. }
	\label{fig:correlation}
\end{figure}

Pearson's correlation coefficient is aimed at detecting linear relationships between two continuous random variables (those whose answer is not categorical (yes, no; present, absent; red, green, blue) or a rank (none, little, medium, large, huge)). It is important that the two variables are random, meaning that the experimenter cannot fix any one of them (for instance the relationship between drug dose and its physiological effect cannot be analyzed by Pearson's correlation, because the experimenter normally is setting the dose). The goal of the pilot study is to determine a confidence interval for the correlation coefficient. The correlation coefficient is a number between -1 and 1, whose absolute value determines the strength of the correlation (0=there is no linear correlation, 1=the two variables are perfectly (positively or negatively) correlated). Qualifying the strength of the correlation coefficient largely depends on the scientific domain (a correlation of 0.5 may be large in one domain and small in another). In the biomedical domain, there is no accepted nomenclature for the different strengths. If we borrow it from behavioral science \citep{Cohen1988}[p.79], we may refer to correlations whose absolute value is below 0.1 as nonexistent, between 0.1 and 0.3 as small, between 0.3 and 0.5 as moderate, and above 0.5 as large. However, these are mere labels and the importance of the correlation in each scientific problem has to be ultimately determined by the researcher. Figure \ref{fig:correlation} shows datasets with different correlation coefficients to illustrate these limits.

As in the case of proportions, we need to assume an expected observed correlation, $r$, and the confidence interval for the correlation is (see the Supplementary Material for details)
\[
   \rho\in\left[
		\mathrm{tanh}\left(\mathrm{tanh}^{-1}(r)-z_{1-\frac{\alpha}{2}}\frac{1}{\sqrt{N-3}}\right),
		\mathrm{tanh}\left(\mathrm{tanh}^{-1}(r)+z_{1-\frac{\alpha}{2}}\frac{1}{\sqrt{N-3}}\right)
	 \right]
\]
From this confidence interval and the desired accuracy, $\delta$, defined as the total width of the confidence interval, we may calculate the number of samples needed. Supplementary Table \ref{tab:correlationSampleSize} shows the sample size for some common values of $\delta$ and $\alpha$. Conversely, we may calculate the accuracy for some typical pilot sample sizes (Supplementary Table \ref{tab:correlationAccuracy}).

\underline{Worked example}:

Let us assume we are interested in studying what we suspect is a moderate correlation between the expression level of a given gene and blood pressure. We expect the correlation to be about 0.3. If we carry out a standard pilot study with about 20 animals, then the 95\% confidence interval would be $[-0.16,0.66]$ (see Supplementary Table \ref{tab:correlationAccuracy}). That is, it goes from a small negative correlation ($\rho=-0.16$), through no effect ($\rho=0$), to a relatively strong positive correlation ($\rho=0.66$). The consequence is that our pilot study is not too clarifying. If we want to construct a confidence interval of a width of $\delta=0.2$, we would need to analyze at least 320 animals (see Supplementary Table \ref{tab:correlationSampleSize}).

\subsection*{Pilot study to determine a lifetime}

\underline{Typical scientific questions}:
\begin{enumerate}
	\item Which is the mean survival time of mice infected with this bacterial strain?
	\item Which is the expected time until animals in an animal facility are not useful for experimentation by contamination by some other contagious disease?
	\item Which is the expected time until an animal gets infected if it is introduced in an infected animal facility?
	\item Which is the mean time until animals in an experiment explore for the first time a given room of a maze?
\end{enumerate}

Some pilot studies try to determine the mean time until an event occurs (being infected, contaminated, or visiting a room). In general, this kind of experiments falls within the statistical domain of ``survival analysis''. The calculation of the sample size when the mean lifetime is totally unknown relies on the assumption of an exponential distribution of the survival time. This is a relatively reasonable assumption when the probability of the event (infection, contamination, ...) is constant over time (technically, its \textit{hazard function} is constant over time). There are situations in which the probability at the beginning is larger and it fades away as time passes by (e.g., the infant mortality is larger than adult mortality), or just the opposite, the probability as time passes by is increased (e.g., the probability of heart failure is larger at old age than at a younger age; or the probability of being infected in an infected colony increases over time). In these two situations, the statistical design based on a constant hazard is incorrect. However, the constant hazard design still provides an order of magnitude of the number of animals required.

Let us refer to the observed mean lifetime as $\hat{\theta}$. As usual, the number of animals required for the experiment depends on the desired precision $\delta=k\hat{\theta}$. We will carry out the experiment (observing the animals until the event occurs) with $N$ animals until we observe $E$ events amongst the $N$ animals. At this time, we will stop the experiment and construct the confidence interval for the mean lifetime. At that moment, there will remain some animals for which we have not observed the event, these animals are said to be censored, and the proportion of censored animals will be referred to as $C$ (for instance, $C=20\%$ means that after observing $E$ events, there still remain 20\% of the $N$ animals for which we have not seen the event).

Let us refer to the true underlying mean lifetime as $\theta$ and to the observed one as $\hat{\theta}$. The confidence interval for the mean lifetime (see Supplementary Material for the detailed derivation)
\[\theta \in \left[\frac{2E\hat{\theta}}{\chi^2_{1-\frac{\alpha}{2},2E}}, \frac{2E\hat{\theta}}{\chi^2_{\frac{\alpha}{2},2E}}\right]\]
If this confidence interval is to be smaller than $\delta=k\hat{\theta}$, then we can calculate the number of required events, $E$ (see corresponding tables in the Supplementary Material). Once we have the required number of events, we can calculate the required number of samples by setting the desired proportion of censored animals, $C$,
\[N=\frac{E}{1-C}\]
Note that $C$ is a parameter that we fix at the time of the experiment design. If $C=20$\%, we will stop the experiment when 20\% of the animals are still ``alive'' (the event did not occur). In this way, we can avoid unnecessarily long waits related to the long tail of the exponential distribution. 

Supplementary Table \ref{tab:expSampleSize} shows the required number of events, $E$, given the desired level of precision and confidence level. Alternatively, Supplementary Table \ref{tab:expAccuracy} shows the accuracy achieved for typical sample sizes used in pilot studies. 

\underline{Worked example}:

Let us assume that we want to determine the mean time of some animals to get infected after exposure to a bacterial strain with a confidence level of 95\% and a precision of $k=20\%$ of the observed mean time. Looking at Supplementary Table \ref{tab:expSampleSize}, we require $E=388$ events. If we use $N=388$ animals for the experiment, the last animals may take really a long time to be infected. Since we would not like to wait that long, we will use more animals so that we stop the experiment when $90\%$ of them are infected. That means that the censored proportion is $C=100\%-90\%=10\%$. That is the number of animals required is $N=\frac{388}{90\%}=432$ animals.

If we use a typical pilot size, like $N=20$, and wait for all animals to be infected (that is, $C=0$, implying $N=E$), then the true mean lifetime lies in the confidence interval $[0.67,1.64]$ (see Supplementary Table \ref{tab:expAccuracy}). That is, the true mean lifetime is, with probability 95\%, between 67\% and 164\% of the observed mean lifetime.

Interestingly, given the mean lifetime, we may estimate the probability per unit of time of the event. For example, assume that the mean lifetime in the infection example is $\theta=5$ days, then everyday the probability of getting infected is $p=\frac{1}{\theta}=1/5=20\%$. Analogously to the mean lifetime, we could construct a confidence interval for the daily probability of infection using the upper and lower bounds of the confidence interval for the mean lifetime.

\subsection*{Mathematical derivations}

\underline{Statistical derivation for standard deviations:}

If we assume that the observations are gaussianly distributed, then we know that the statistic
\[\chi^2=\frac{(N-1)s^2}{\sigma_0^2}\]
is distributed with a $\chi^2$ distribution and $N-1$ degrees of freedom, where $N$ is the number of independent samples used in the experiment, $s^2$ is the sample variance and $\sigma_0^2$ is the true underlying variance. Then, we may construct a confidence interval with confidence $1-\alpha$ exploiting the fact that
	\[\begin{array}{c}
		\mathrm{Pr}\left\{\chi^2_{\frac{\alpha}{2},N-1}\leq \frac{(N-1)s^2}{\sigma_0^2} \leq \chi^2_{1-\frac{\alpha}{2},N-1} \right\}=1-\alpha \\
	\end{array}\]
	That is, with probability $1-\alpha$, the true standard deviation is within the range
	\[\begin{array}{c}
		 s\sqrt{\frac{N-1}{\chi^2_{1- \frac{\alpha}{2},N-1}}} \leq \sigma_0 \leq s\sqrt{\frac{N-1}{\chi^2_{\frac{\alpha}{2},N-1}}}
	\end{array}\]
	The righthand side is more restrictive than the lefthand side, so that if the precision is $\delta$, $N$ must be such that
	\begin{equation}\sqrt{\frac{N-1}{\chi^2_{\frac{\alpha}{2},N-1}}}\leq 1+\delta\label{eq:stdDesign}\end{equation}
	In other words, we must find the smallest integer $N$ such that the inequality above is fulfilled. 
	
\underline{Statistical derivation for means:}

Let us assume that our goal is to determine the mean value of a variable (for instance, the gene expression level of a given gene under specific conditions). Let us construct the statistic
\[t=\frac{\mu-\hat{\mu}}{s}\]
that determines the size of the estimation error (that is, the difference between the true mean $\mu$ and our estimate $\hat{\mu}$) with respect to the sample standard deviation, $s$. If the observed data is normally distributed, then we know that $\sqrt{N}t$ has a Student's $t$ distribution with $N-1$ degrees of freedom (being $N$ the number of measurements in our experiment) and that
\[\mathrm{Pr}\left\{\left|\frac{\mu-\hat{\mu}}{s/\sqrt{N}}\right|\leq t_{1-\frac{\alpha}{2},N-1}\right\}=1-\alpha\]
That is, with probability $1-\alpha$, the true mean is within the range
\[\hat{\mu}-t_{1-\frac{\alpha}{2},N-1}\frac{s}{\sqrt{N}}\leq \mu \leq \hat{\mu}+t_{1-\frac{\alpha}{2},N-1}\frac{s}{\sqrt{N}}\]
If the desired precision is $|t|=\left|\frac{\mu-\hat{\mu}}{s}\right|\leq\delta$, then, $N$ must be such that
\begin{equation}\frac{t_{1-\frac{\alpha}{2},N-1}}{\sqrt{N}}\leq \delta\label{eq:meanDesign}\end{equation}

\underline{Statistical derivation for proportions between 10\% and 90\%:}

The correct experimental design involves calculations with the binomial distribution. It can be shown \citep{Fleiss2003}[p. 25] that if $r$ cases are observed in a set of $N$ animals, then the exact two-sided confidence interval is (Clopper-Pearson formula)
\[\frac{r}{r+(N-r+1)F_{1-\frac{\alpha}{2};2(N-r+1),2r}} \leq p \leq \frac{(r+1)F_{1-\frac{\alpha}{2};2(r+1),2(N-r)}}{(N-r)+(r+1)F_{1-\frac{\alpha}{2};2(r+1),2(N-r)}}\]
where $F_{y;m,n}$ is the $y$-th percentile of Snedecor's F with $m$ and $n$ degrees of freedom. As can be seen, the exact expression does not easily lend itself to the calculation of the sample size. 
If the true proportion is $p$ and $N$ is the number of animals in the experiment, the binomial distribution can be asymptotically approximated by a normal distribution if $Np>5$ and $N(1-p)>5$ (for $N<50$, these conditions are usually met for $0.1\leq p\leq 0.9$; the typical sample size for a pilot study is normally below $N<20$, which would further restrain $p$ to be between 25\% and 75\%). If the Gaussian approximation holds, then we may construct the $1-\alpha$ confidence interval exploiting the fact that
\[\mathrm{Pr}\left\{\left|\frac{p-\hat{p}}{\sqrt{\frac{\hat{p}(1-\hat{p})}{N}}}\right|\leq z_{1-\frac{\alpha}{2}}\right\}=1-\alpha\]
That is, with probability $1-\alpha$, the true proportion is in the interval
\[\hat{p}-z_{1-\frac{\alpha}{2}}\sqrt{\frac{\hat{p}(1-\hat{p})}{N}}\leq p \leq \hat{p}+z_{1-\frac{\alpha}{2}}\sqrt{\frac{\hat{p}(1-\hat{p})}{N}}\]
If the desired precision is $|p-\hat{p}|\leq \delta$, then, $N$ must be such that
\begin{equation} z_{1-\frac{\alpha}{2}}\sqrt{\frac{\hat{p}(1-\hat{p})}{N}}\leq \delta \end{equation}
or 
\begin{equation} z_{1-\frac{\alpha}{2}}\sqrt{\frac{\hat{p}(1-\hat{p})}{N}}+\frac{1}{2N}\leq \delta \end{equation}
if we apply the so-called continuity correction (a correction that accounts for the fact that the Gaussian is a continuous distribution while the binomial is a discrete one).

As can be seen, this design formula differs from the previous ones (Eqs. (\ref{eq:stdDesign}) and (\ref{eq:meanDesign})) in that $N$ depends on the observed value $\hat{p}$ (note that before performing the experiment, $\hat{p}$ is unknown and the experimenter must make some assumption about its expected value; the worse case is for $\hat{p}=50\%$ and in those experiments in which nothing is known about the possible experiment outcome, this is the value that must be used for the experiment design). 

\ul{Statistical derivation for two-sided intervals of extreme proportions ($p$<10\% or $p$>90\%):}

The exact Clopper-Pearson formula of the confidence interval in the previous section is still valid. However, as we have seen above, it does not lend itself to easy calculations by hand. For doing so, we present here an approximate formula using the Gaussian. Let $p$ be the small proportion ($0<p<10\%$) we are trying to estimate. If we observe a total of $N$ individuals, the number $X$ of observed individuals with the property sought (the rare disease or the adverse effect) follows a Binomial distribution with parameters $N$ and $p$ ($B(N,p)$). If $N\geq 20$ and $p\leq 5\%$, then the Binomial distribution can be safely approximated by a Poisson distribution with parameter $Np$. If additionally $Np\geq 30$, then the Poisson distribution can be approximated by a Gaussian whose mean is $Np$ and its variance is also $Np$. That means that the confidence interval can be constructed knowing that
\[\mathrm{Pr}\left\{\left|\frac{X-Np}{\sqrt{Np}}\right|\leq z_{1-\frac{\alpha}{2}}\right\}=
  \mathrm{Pr}\left\{\left|\frac{\frac{X}{N}-p}{\sqrt{\frac{p}{N}}}\right|\leq z_{1-\frac{\alpha}{2}}\right\}1-\alpha\]
That is, with probability $1-\alpha$, the true mean is within the range
\[\hat{p}-z_{1-\frac{\alpha}{2}}\sqrt{\frac{p}{N}}\leq p \leq \hat{p}+z_{1-\frac{\alpha}{2}}\sqrt{\frac{p}{N}}\]
Then, the precision is 
\[z_{1-\frac{\alpha}{2}}\sqrt{\frac{p}{N}}\leq \delta=kp\]
From where we can solve for the number of individuals for our experiment
\[N\geq \left(\frac{z_{1-\frac{\alpha}{2}}}{k\sqrt{p}}\right)^2\]

\ul{Statistical derivation for one-sided intervals of extreme proportions ($p$<10\% or $p$>90\%):}

The confidence interval constructed in this case is one-sided, since we are interested in showing that the underlying proportion is between 0 and a upper bound, $p_U$ (for instance, [0,1]\% in the first example of the corresponding section in the main text) or between a lower bound, $p_L$, and 100\% (for instance, [99,100]\% in the last example). These two values determine the accuracy of our result. The exact binomial formulas (Clopper-Pearson) guarantees that with probability $1-\alpha$, the true underlying proportion is between
\[0 \leq p \leq \frac{(r+1)F_{1-\alpha;2(r+1),2(N-r)}}{(N-r)+(r+1)F_{1-\alpha;2(r+1),2(N-r)}}\]
if $p$ is low, or
\[\frac{r}{r+(N-r+1)F_{1-\alpha;2(N-r+1),2r}} \leq p \leq 1\]
if $p$ is high.

As we did above, this formula is inconvenient for manual calculations. Moreover, the formula allows for several statistical successes, $r$, amongst the $N$ observed animals, which makes it more difficult to understand. We may restrict more the experiment. We will analyze $N$ animals, if none of them has the property that has probability $p$, then we accept the alternative hypothesis (in Example 1, if no mouse is infected after being vaccinated, then we accept the hypothesis that the probability of being infected after vaccination is smaller than 1\%). From the binomial distribution, we know that the probability of observing 0 events is given by
\[\mathrm{Pr}\{X=0\}=\left(\begin{array}{c}N \\ 0\end{array}\right)p_U^0(1-p_U)^{N-0}=(1-p_U)^N\]
If the null hypothesis is true, then it must be
\[(1-p_U)^N\geq \alpha\]
From which it is easy to derive
\[N\geq \frac{\log(\alpha)}{\log(1-p_U)}\]
if the probability $p$ is low or
\[N\geq \frac{\log(\alpha)}{\log(p_L)}\]
if $p$ is high. This strategy is called zero-acceptance experiment and it is the standard procedure for instance for disease controls in animal facilities.

This formula can be easily calculated. However, in Epidemiology there is a widespread approximation known as the \textit{rule of three}. The random variable $2Np$ is distributed as a $\chi^2$ distribution with $2(r+1)$ degrees of freedom (the same rule can be derived following a reasoning from the Poisson distribution). For $r=0$ we must have
\[2Np_U\geq \chi^2_{1-\alpha;2} \Rightarrow N\geq \frac{\chi^2_{1-\alpha;2}}{2p_U}\]
or alternatively
\[2N(1-p_L)\geq \chi^2_{1-\alpha;2} \Rightarrow N\geq \frac{\chi^2_{1-\alpha;2}}{2(1-p_L)}\]
For $\alpha=0.05$ it is
\[N\geq\frac{2.9957}{p_U}\]
which is the well-known rule of three. Remind that this rule was derived under the assumption that we will observe $N$ animals, if none of them is infected (fulfills the property), then we accept the alternative hypothesis.

\underline{Statistical derivation for correlations}:

If the the two variables being correlated are normally distributed, and the experimentally observed correlation is $r$, then the statistic
\[z_r=\mathrm{tanh}^{-1}(r)=\frac{1}{2}\log\frac{1+r}{1-r}\]
follows a normal distribution with mean $\mathrm{tanh}^{-1}(\rho)$ (where $\mathrm{tanh}^{-1}$ is the hyperbolic arc tangent and $\rho$ is the true underlying correlation) and standard deviation $\frac{1}{\sqrt{N-3}}$. That is, the two-sided confidence interval is given by
\[\mathrm{Pr}\left\{\left|\frac{\mathrm{tanh}^{-1}(r)-\mathrm{tanh}^{-1}(\rho)}{\frac{1}{\sqrt{N-3}}}\right|<z_{1-\frac{\alpha}{2}} \right\}=1-\alpha\]
That is with probability $1-\alpha$, the true correlation coefficient is in the range
\[\mathrm{tanh}^{-1}(r)-z_{1-\frac{\alpha}{2}}\frac{1}{\sqrt{N-3}}<\mathrm{tanh}^{-1}(\rho)<\mathrm{tanh}^{-1}(r)-z_{1-\frac{\alpha}{2}}\frac{1}{\sqrt{N-3}}\]
A remarkable property is that this range is not symmetric about the observed correlation (it is symmetric about its hyperbolic tangent). To design a pilot experiment, we need to provide an expected level of correlation, $\rho$, and the $N$ is chosen so that the interval between the upper and lower bounds is smaller than a given precision $\delta$
being
\begin{equation}
   \mathrm{tanh}\left(\mathrm{tanh}^{-1}(\rho)+z_{1-\frac{\alpha}{2}}\frac{1}{\sqrt{N-3}}\right)-\mathrm{tanh}\left(\mathrm{tanh}^{-1}(\rho)-z_{1-\frac{\alpha}{2}}\frac{1}{\sqrt{N-3}}\right)
	\leq \delta
\end{equation}

\underline{Statistical derivation for lifetimes}:

The maximum likelihood estimate of the mean lifetime is \citep{Mathews2010}[Chap. 9]
\[\hat{\theta}=\frac{1}{E}\sum\limits_{i=1}^Nt_i\]
where $t_i$ is the time at which the event occurred for the $i$-th animal or the maximum time of observation (i.e., the time at which we have observed $E$ events in total, this is called Type II Censoring). The random variable $\frac{2E\hat{\theta}}{\theta}$ is distributed as a $\chi^2$ variable with $2E$ degrees of freedom. Then
\[\mathrm{Pr}\left\{\chi^2_{\frac{\alpha}{2},2E}\leq \frac{2E\hat{\theta}}{\theta} \leq \chi^2_{1-\frac{\alpha}{2},2E}\right\}=1-\alpha\]
That is, with probability $1-\alpha$, the true mean lifetime is in the interval
\[\frac{2E\hat{\theta}}{\chi^2_{1-\frac{\alpha}{2},2E}} \leq \theta \leq \frac{2E\hat{\theta}}{\chi^2_{\frac{\alpha}{2},2E}}\]
If this confidence interval is to be smaller than $\delta=k\hat{\theta}$, then the number of events must fulfill
\[2E\left(\frac{1}{\chi^2_{\frac{\alpha}{2},2E}}-\frac{1}{\chi^2_{1-\frac{\alpha}{2},2E}}\right)\leq k\]
There is no closed form solution for $E$ and a numerical method is required. Once we solve for the number of events, the number of required animals is
\[N=\frac{E}{1-C}\]

\newpage

\begin{landscape}

\begin{table}[htbp]
	\centering
	\begin{tabular}{|c|c|c|c|c|c|c|c|}
		\hline
			 & & \multicolumn{6}{c|}{$\delta$ (Accuracy)} \\
		\hline
			 & & 1\% & 5\% & 10\% & 20\% & 50\% & 100\% \\
		\hline
			\multirow{3}{*}{$1-\alpha$ (Confidence)}& 90\% & 13823 & 602 & 167 & 51 & 14 & 7\\
			                                        & 95\% & 19597 & 849 & 234 & 70 & 18 & 9 \\
			                                        & 99\% & 33798 & 1455 & 398 & 118 & 29 & 13 \\
		\hline
	\end{tabular}
	\caption{Sample size for the determination of the standard deviation given the accuracy ($\delta$) and confidence level ($1-\alpha$).}
	\label{tab:varianceSampleSize}
\end{table}	

\begin{table}[htbp]
	\centering
	\begin{tabular}{|c|c|c|c|c|c|c|c|}
		\hline
			 & & \multicolumn{6}{c|}{$N$ (Sample size)} \\
		\hline
			 & & 5 & 10 & 15 & 20 & 25 & 30 \\
		\hline
			\multirow{3}{*}{$1-\alpha$ (Confidence)}& 90\% & 137.24\% & 64.52\% & 45.97\% & 37.04\% & 31.65\% & 27.97\%\\
			                                        & 95\% & 187.36\% & 82.56\% & 57.71\% & 46.06\% & 39.12\% & 34.43\% \\
			                                        & 99\% & 339.60\% & 127.76\% & 85.36\% & 66.62\% &  55.81\% & 48.67\% \\
		\hline
	\end{tabular}
	\caption{Accuracy ($\delta$) of the standard deviation estimate given the sample size ($N$) and confidence level ($1-\alpha$).}
	\label{tab:varianceAccuracy}
\end{table}	

\end{landscape}
\begin{landscape}

\begin{table}[htbp]
	\centering
	\begin{tabular}{|c|c|c|c|c|c|c|c|}
		\hline
			 & & \multicolumn{6}{c|}{$\delta$ (Accuracy)} \\
		\hline
			 & & 1\% & 5\% & 10\% & 20\% & 50\% & 100\% \\
		\hline
			\multirow{3}{*}{$1-\alpha$ (Confidence)}& 90\% & 27058 & 1085 & 273 & 70 & 13 & 5 \\
			                                        & 95\% & 38418 & 1540 & 387 & 99 & 18 & 7 \\
			                                        & 99\% & 66353 & 2658 & 668 & 170 & 31 & 11 \\
		\hline
	\end{tabular}
	\caption{Sample size for the determination of the mean given the accuracy ($\delta$) and confidence level ($1-\alpha$).}
	\label{tab:meanSampleSize}
\end{table}	

\begin{table}[htbp]
	\centering
	\begin{tabular}{|c|c|c|c|c|c|c|c|}
		\hline
			 & & \multicolumn{6}{c|}{$N$ (Sample size)} \\
		\hline
			 & & 5 & 10 & 15 & 20 & 25 & 30 \\
		\hline
			\multirow{3}{*}{$1-\alpha$ (Confidence)}& 90\% & 95.33\% & 57.97\% & 45.48\% & 38.66\% & 34.22\% & 31.02\% \\
			                                        & 95\% & 124.17\%& 71.54\% & 55.38\% & 46.80\% & 41.28\% & 37.34\%\\
			                                        & 99\% & 205.90\% & 102.77\% & 76.86\% & 63.97\% & 55.94\% & 50.32\% \\
		\hline
	\end{tabular}
	\caption{Accuracy of the mean estimate given the sample size ($N$) and confidence level ($1-\alpha$).}
	\label{tab:meanAccuracy}
\end{table}	

\end{landscape}

\begin{table}[htbp]
	\centering
	\hspace*{-1cm}\begin{tabular}{|c|c|c|c|c|c|c|c|}
		\hline
			 & & \multicolumn{6}{c|}{$\delta$ (Accuracy)} \\
		\hline
			 & & 1\% & 5\% & 10\% & 15\% & 20\% & 25\% \\
		\hline
			 & & \multicolumn{6}{c|}{$\hat{p}=10\%$ or $\hat{p}=90\%$} \\
		\hline
			\multirow{3}{*}{$1-\alpha$ (Confidence)}& 90\% & 2534 (2535) & 117 (117) & 33 (\sout{34}) & 16 (\sout{11}) & 10  (\sout{6}) &  7 (\sout{4}) \\
			                                        & 95\% & 3557 (3557) & 158 (158) & 44 (\sout{45}) & 21 (\sout{14}) & 13  (\sout{7}) &  8 (\sout{5}) \\
			                                        & 99\% & 6071 (6071) & 258 (259) & 69 (70)        & 32 (\sout{20}) & 19 (\sout{10}) & 12 (\sout{6}) \\
		\hline
			 & & \multicolumn{6}{c|}{$\hat{p}=15\%$ or $\hat{p}=85\%$} \\
		\hline
			\multirow{3}{*}{$1-\alpha$ (Confidence)}& 90\% & 3548 (3549) & 157 (158) & 43 (44) & 21 (\sout{22}) & 12 (\sout{10}) &  8 (\sout{6}) \\
			                                        & 95\% & 4997 (4998) & 215 (216) & 58 (59) & 27 (\sout{29}) & 16 (\sout{12}) & 10 (\sout{7}) \\
			                                        & 99\% & 8558 (8560) & 356 (359) & 93 (95) & 42 (45)        & 24 (\sout{18}) & 15 (\sout{10}) \\
		\hline
			 & & \multicolumn{6}{c|}{$\hat{p}=20\%$ or $\hat{p}=80\%$} \\
		\hline
			\multirow{3}{*}{$1-\alpha$ (Confidence)}& 90\% &  4428  (4429) & 192 (193) &  52  (53) & 24 (26)        & 14 (\sout{16}) &  9 (\sout{8}) \\
			                                        & 95\% &  6245  (6246) & 264 (266) &  70  (72) & 32 (34)        & 18 (\sout{21}) & 12 (\sout{10}) \\
			                                        & 99\% & 10713 (10716) & 442 (445) & 114 (116) & 51 (54)        & 29 (32)        & 18 (\sout{15}) \\
		\hline
			 & & \multicolumn{6}{c|}{$\hat{p}=25\%$ or $\hat{p}=75\%$} \\
		\hline
			\multirow{3}{*}{$1-\alpha$ (Confidence)}& 90\% &  5171  (5173) & 221 (223) &  59  (61) & 28 (29) & 16 (\sout{18}) & 10 (\sout{12}) \\
			                                        & 95\% &  7301  (7303) & 306 (308) &  80  (82) & 37 (39) & 21 (23)        & 13 (\sout{16}) \\
			                                        & 99\% & 12538 (12541) & 515 (518) & 131 (135) & 59 (62) & 33 (36)        & 21 (24) \\
		\hline
			 & & \multicolumn{6}{c|}{$\hat{p}=30\%$ or $\hat{p}=70\%$} \\
		\hline
			\multirow{3}{*}{$1-\alpha$ (Confidence)}& 90\% &  5780  (5782) & 246 (247) &  65  (67) & 30 (32) & 17 (19)        & 11 (\sout{13}) \\
			                                        & 95\% &  8165  (8167) & 341 (343) &  89  (91) & 40 (43) & 23 (25)        & 15 (17) \\
			                                        & 99\% & 14030 (14034) & 574 (578) & 146 (150) & 65 (69) & 36 (40)        & 23 (27) \\
		\hline
			 & & \multicolumn{6}{c|}{$\hat{p}=35\%$ or $\hat{p}=65\%$} \\
		\hline
			\multirow{3}{*}{$1-\alpha$ (Confidence)}& 90\% &  6253  (6255) & 265 (266) &  70  (72) &  32 (34) &  19 (21) & 12 (\sout{14}) \\
			                                        & 95\% &  8837  (8840) & 367 (370) &  95  (98) &  43 (46) &  25 (27) & 16 (18) \\
			                                        & 99\% & 15191 (15195) & 620 (624) & 158 (161) &  70 (74) &  39 (43) & 25 (29) \\
		\hline
			 & & \multicolumn{6}{c|}{$\hat{p}=40\%$ or $\hat{p}=60\%$} \\
		\hline
			\multirow{3}{*}{$1-\alpha$ (Confidence)}& 90\% &  6592  (6593) &  278 (280) &  73  (75) & 34 (36) & 19 (21) & 13 (15) \\
			                                        & 95\% &  9317  (9320) &  387 (389) & 100 (102) & 45 (48) & 26 (28) & 16 (19) \\
			                                        & 99\% & 16020 (16024) &  654 (657) & 166 (170) & 74 (78) & 41 (45) & 26 (30) \\
		\hline
			 & & \multicolumn{6}{c|}{$\hat{p}=45\%$ or $\hat{p}=55\%$} \\
		\hline
			\multirow{3}{*}{$1-\alpha$ (Confidence)}& 90\% &  6795  (6796) &  286 (288) &  75  (77) & 35 (37) & 20 (22) & 13 (15) \\
			                                        & 95\% &  9605  (9608) &  398 (401) & 103 (105) & 47 (49) & 26 (29) & 17 (20) \\
			                                        & 99\% & 16518 (16522) &  673 (677) & 171 (175) & 76 (80) & 42 (46) & 27 (31) \\
		\hline
			 & & \multicolumn{6}{c|}{$\hat{p}=50\%$} \\
		\hline
			\multirow{3}{*}{$1-\alpha$ (Confidence)}& 90\% &  6862  (6864) &  289 (291) &  76  (78) & 35 (37) & 20 (22) & 13 (15) \\
			                                        & 95\% &  9701  (9704) &  402 (404) & 104 (106) & 47 (50) & 27 (29) & 17 (20) \\
			                                        & 99\% & 16684 (16688) &  680 (684) & 172 (176) & 77 (81) & 43 (47) & 27 (31) \\
		\hline
	\end{tabular}
	\caption{Sample size for the determination of a proportion given the accuracy ($\delta$) and confidence level ($1-\alpha$). The number without parenthesis
	   is the sample size calculated through the exact binomial formula, and in parenthesis the one calculated by the Gaussian approximation with continuity correction
		 (we have crossed out those sample
	   sizes for which the Gaussian approximation does not hold).}
	\label{tab:proportionSampleSize}
\end{table}	

\begin{table}[htbp]
	\centering
	\hspace*{-1cm}\begin{tabular}{|c|c|c|c|c|c|c|c|}
		\hline
			 & & \multicolumn{6}{c|}{$N$ (Sample size)} \\
		\hline
			 & & 5 & 10 & 15 & 20 & 25 & 30 \\
		\hline
			 & & \multicolumn{6}{c|}{ $\hat{p}=10\%$} \\
		\hline
			\multirow{3}{*}{$1-\alpha$ (Confidence)}& 90\% & [0,56.3] & [0.5,39.4] & [1.2,32.3] & [1.8,28.3] & [2.3,25.7] & [2.8,23.9] \\
			                                        & 95\% & [0,62.9] & [0.3,44.5] & [0.7,36.3] & [1.2,31.7] & [1.7,28.7] & [2.1,26.5] \\
			                                        & 99\% & [0,74.4] & [0.1,54.4] & [0.2,44.6] & [0.5,38.7] & [0.8,34.8] & [1.2,32.0] \\
		\hline
			 & & \multicolumn{6}{c|}{ $\hat{p}=15\%$} \\
		\hline
			\multirow{3}{*}{$1-\alpha$ (Confidence)}& 90\% & [0.3,61.2] & [1.8,45.2] & [3.2,38.3] & [4.2,34.4] & [5.1,31.8] & [5.7,29.9] \\
			                                        & 95\% & [0.1,67.4] & [1.1,50.3] & [2.2,42,4] & [3.2,37.9] & [4.0,34.9] & [4.7,32.7] \\
			                                        & 99\% & [0.0,78.1] & [0.4,59.9] & [1.0,50.6] & [1.8,44.9] & [2.4,41.2] & [3.0,38.4] \\
		\hline
			 & & \multicolumn{6}{c|}{ $\hat{p}=20\%$} \\
		\hline
			\multirow{3}{*}{$1-\alpha$ (Confidence)}& 90\% & [1.0,65.7] & [3.7,50.7] & [5.7,44.0] & [7.1,40.1] & [8.2,37.5] & [9.1,35.7] \\
			                                        & 95\% & [0.5,71.6] & [2.5,55.6] & [4.3,48.1] & [5.7,43.7] & [6.8,40.7] & [7.7,38.6] \\
			                                        & 99\% & [0.1,81.5] & [1.1,64.8] & [2.4,56.1] & [3.6,50.7] & [4.6,47.0] & [5.4,44.3] \\
		\hline
			 & & \multicolumn{6}{c|}{ $\hat{p}=25\%$} \\
		\hline
			\multirow{3}{*}{$1-\alpha$ (Confidence)}& 90\% & [2.1,70.0] & [6.0,55.8] & [8.6,49.3] & [10.4,45.6] & [11.7,43.0] & [12.7,41.2] \\
			                                        & 95\% & [1.2,75.5] & [4.4,60.6] & [6.9,53.4] & [ 8.7,49.1] & [10.0,46.2] & [11.1,44.1] \\
			                                        & 99\% & [0.3,84.5] & [2.2,69.3] & [4.2,61.1] & [ 5.8,56.0] & [ 7.2,52.4] & [ 8.3,49.8] \\
		\hline
			 & & \multicolumn{6}{c|}{ $\hat{p}=30\%$} \\
		\hline
			\multirow{3}{*}{$1-\alpha$ (Confidence)}& 90\% & [3.6,73.9] & [8.7,60.7] & [11.9,54.5] & [14.0,50.8] & [15.5,48.3] & [16.6,46.5] \\
			                                        & 95\% & [2.3,79.1] & [6.7,65.2] &  [9.7,58.4] & [11.9,54.3] & [13.5,51.5] & [14.7,49.4] \\
			                                        & 99\% & [0.8,87.2] & [3.7,73.5] &  [6.4,65.8] &  [8.5,61.0] & [10.1,57.5] & [11.4,55.0] \\
		\hline
			 & & \multicolumn{6}{c|}{ $\hat{p}=35\%$} \\
		\hline
			\multirow{3}{*}{$1-\alpha$ (Confidence)}& 90\% & [5.5,77.6] & [11.7,65.3] & [15.4,59.4] & [17.7,55.8] & [19.4,53.4] & [20.7,51.6] \\
			                                        & 95\% & [3.6,82.3] &  [9.3,69.6] & [12.9,63.2] & [15.4,59.2] & [17.2,56.5] & [18.6,54.5] \\
			                                        & 99\% & [1.4,89.6] &  [5.5,77.4] &  [8.9,70.3] & [11.4,65.7] & [13.3,62.4] & [14.8,60.0] \\
		\hline
			 & & \multicolumn{6}{c|}{ $\hat{p}=40\%$} \\
		\hline
			\multirow{3}{*}{$1-\alpha$ (Confidence)}& 90\% & [7.6,81.1] & [15.0,69.6] & [19.1,64.0] & [21.7,60.6] & [23.6,58.3] & [25.0,56.6] \\
			                                        & 95\% & [5.3,85.3] & [12.2,73.8] & [16.3,67.7] & [19.1,63.9] & [21.1,61.3] & [22.7,59.4] \\
			                                        & 99\% & [2.3,91.7] &  [7.7,80.9] & [11.7,74.4] & [14.6,70.1] & [16.8,67.0] & [18.5,64.7] \\
		\hline
			 & & \multicolumn{6}{c|}{ $\hat{p}=45\%$} \\
		\hline
			\multirow{3}{*}{$1-\alpha$ (Confidence)}& 90\% & [10.1,84.3] & [18.5,73.8] & [23.0,68.5] & [25.9,65.3] & [27.9,63.1] & [29.4,61.4] \\
			                                        & 95\% &  [7.2,88.1] & [15.3,77.6] & [20.0,72.0] & [23.1,68.5] & [25.2,66.0] & [26.9,64.1] \\
			                                        & 99\% &  [3.4,93.6] & [10.1,84.2] & [14.8,78.3] & [18.1,74.3] & [20.5,71.4] & [22.4,69.2] \\
		\hline
			 & & \multicolumn{6}{c|}{ $\hat{p}=50\%$} \\
		\hline
			\multirow{3}{*}{$1-\alpha$ (Confidence)}& 90\% & [12.8,87.2] & [22.2,77.8] & [27.1,72.9] & [30.2,69.8] & [32.3,67.7] & [33.9,66.1] \\
			                                        & 95\% &  [9.4,90.6] & [18.7,81.3] & [23.9,76.1] & [27.2,72.8] & [29.5,70.5] & [31.3,68.7] \\
			                                        & 99\% &  [4.8,95.2] & [12.8,87.2] & [18.1,81.9] & [21.8,78.2] & [24.4,75.6] & [26.5,73.5] \\
		\hline
	\end{tabular}
	\caption{Proportion confidence interval (in \%) given the sample size ($N$) and confidence level ($1-\alpha$). Because of the low $N$, the Gaussian approximation does not hold for most of the cells and these confidence intervals have been calculated using the exact Clopper-Pearson formula (note that the confidence interval does not need to be symmetric).}
	\label{tab:proportionAccuracy}
\end{table}	

\begin{table}[htbp]
	\centering
	\hspace*{-1cm}\begin{tabular}{|c|c|c|c|c|c|c|c|}
		\hline
			 & & \multicolumn{6}{c|}{$N$ (Sample size)} \\
		\hline
			 & & 5 & 10 & 15 & 20 & 25 & 30 \\
		\hline
			 & & \multicolumn{6}{c|}{ $\hat{p}=55\%$} \\
		\hline
			\multirow{3}{*}{$1-\alpha$ (Confidence)}& 90\% & [15.7,89.9] & [26.2,81.5] & [31.5,77.0] & [34.7,74.1] & [36.9,72.1] & [38.6,70.7] \\
			                                        & 95\% & [11.9,92.8] & [22.4,84.7] & [28.0,80.0] & [31.5,76.9] & [34.0,74.8] & [35.9,73.1] \\
			                                        & 99\% &  [6.4,96.6] & [15.8,89.9] & [21.7,85.2] & [25.7,81.9] & [28.6,79.5] & [30.8,77.6] \\
		\hline
			 & & \multicolumn{6}{c|}{ $\hat{p}=60\%$} \\
		\hline
			\multirow{3}{*}{$1-\alpha$ (Confidence)}& 90\% & [18.9,92.4] & [30.4,85.0] & [36.0,80.9] & [39.4,78.3] & [41.7,76.4] & [43.4,75.0] \\
			                                        & 95\% & [14.7,94.7] & [26.2,87.8] & [32.3,83.7] & [36.1,80.9] & [38.7,78.9] & [40.6,77.3] \\
			                                        & 99\% &  [8.3,97.7] & [19.1,92.3] & [25.6,88.3] & [29.9,85.4] & [33.0,83.2] & [35.3,81.5] \\
		\hline
			 & & \multicolumn{6}{c|}{ $\hat{p}=65\%$} \\
		\hline
			\multirow{3}{*}{$1-\alpha$ (Confidence)}& 90\% & [22.4,94.5] & [34.7,88.3] & [40.6,84.6] & [44.2,82.3] & [46.6,80.6] & [48.4,79.3] \\
			                                        & 95\% & [17.7,96.4] & [30.4,90.7] & [36.8,87.1] & [40.8,84.6] & [43.5,82.8] & [45.5,81.4] \\
			                                        & 99\% & [10.4,98.6] & [22.6,94.5] & [29.7,91.1] & [34.3,88.6] & [37.6,86.7] & [40,85.2] \\
		\hline
			 & & \multicolumn{6}{c|}{ $\hat{p}=70\%$} \\
		\hline
			\multirow{3}{*}{$1-\alpha$ (Confidence)}& 90\% & [26.1,96.4] & [39.3,91.3] & [45.5,88.1] & [49.2,86.0] & [51.7,84.5] & [53.5,83.4] \\
			                                        & 95\% & [20.9,97.7] & [34.8,93.3] & [41.6,90.3] & [45.7,88.1] & [48.5,86.5] & [50.6,85.3] \\
			                                        & 99\% & [12.8,99.2] & [26.5,96.3] & [34.2,93.6] & [39.0,91.5] & [42.5,89.9] & [45.0,88.6] \\
		\hline
			 & & \multicolumn{6}{c|}{ $\hat{p}=75\%$} \\
		\hline
			\multirow{3}{*}{$1-\alpha$ (Confidence)}& 90\% & [30.0,97.9] & [44.2,94.0] & [50.7,91.4] & [54.4,89.6] & [57.0,88.3] & [58.8,87.3] \\
			                                        & 95\% & [24.5,98.8] & [39.4,95.6] & [46.6,93.1] & [50.9,91.3] & [53.8,90.0] & [55.9,88.9] \\
			                                        & 99\% & [15.5,99.7] & [30.7,97.8] & [38.9,95.8] & [44.0,94.2] & [47.6,92.8] & [50.2,91.7] \\
		\hline
			 & & \multicolumn{6}{c|}{ $\hat{p}=80\%$} \\
		\hline
			\multirow{3}{*}{$1-\alpha$ (Confidence)}& 90\% & [34.3,99.0] & [49.3,96.3] & [56.0,94.3] & [59.9,92.9] & [62.5,91.8] & [64.3,90.9] \\
			                                        & 95\% & [28.4,99.5] & [44.4,97.5] & [51.9,95.7] & [56.3,94.3] & [59.3,93.2] & [61.4,92.3] \\
			                                        & 99\% & [18.5,99.9] & [35.2,98.9] & [43.9,97.6] & [49.3,96.4] & [53.0,95.4] & [55.7,94.6] \\
		\hline
			 & & \multicolumn{6}{c|}{ $\hat{p}=85\%$} \\
		\hline
			\multirow{3}{*}{$1-\alpha$ (Confidence)}& 90\% & [38.8,99.7] & [54.8,98.2] & [61.7,96.8] & [65.6,95.8] & [68.2,94.9] & [70.1,94.3] \\
			                                        & 95\% & [32.6,99.9] & [49.7,98.9] & [57.6,97.8] & [62.1,96.8] & [65.1,96.0] & [67.3,95.3] \\
			                                        & 99\% &  [21.9,100] & [40.1,99.6] & [49.4,99.0] & [55.1,98.2] & [58.8,97.6] & [61.6,97.0] \\
		\hline
			 & & \multicolumn{6}{c|}{ $\hat{p}=90\%$} \\
		\hline
			\multirow{3}{*}{$1-\alpha$ (Confidence)}& 90\% &  [43.7,100] & [60.6,99.5] & [67.7,98.8] & [71.7,98.2] & [74.3,97.7] & [76.1,97.2] \\
			                                        & 95\% &  [37.1,100] & [55.5,99.7] & [63.7,99.3] & [68.3,98.8] & [71.3,98.3] & [73.5,97.9] \\
			                                        & 99\% &  [25.6,100] & [45.6,99.9] & [55.4,99.8] & [61.3,99.5] & [65.2,99.2] & [68.0,98.8] \\
		\hline
	\end{tabular}
	\caption{Table \ref{tab:proportionAccuracy} continued}
	\label{tab:proportionAccuracyCont}
\end{table}	

\begin{landscape}

\begin{table}[htbp]
	\centering
	\begin{tabular}{|c|c|c|c|c|c|}
		\hline
			 & & \multicolumn{4}{c|}{$k$ (Accuracy)} \\
		\hline
			 & & 10\% & 25\% & 50\% & 100\% \\
		\hline
			 & & \multicolumn{4}{c|}{$p=5\%$ or $p=95\%$} \\
		\hline
			\multirow{3}{*}{$1-\alpha$ (Confidence)}& 90\% & 5342 (5412) & 903 (866) & 246 (\sout{217}) & 72 (\sout{55}) \\
			                                        & 95\% & 7501 (7683) & 1250 (1230) & 334 (\sout{308}) & 94 (\sout{77}) \\
			                                        & 99\% & 12810 (13270) & 2101 (2124) & 548 (\sout{531}) & 149 (\sout{133}) \\
		\hline
			 & & \multicolumn{4}{c|}{$p=2.5\%$ or $p=97.5\%$} \\
		\hline
			\multirow{3}{*}{$1-\alpha$ (Confidence)}& 90\% & 10956 (10823) & 1852 (1732) &  506  (\sout{433}) & 148 (\sout{109}) \\
			                                        & 95\% & 15389 (15366) & 2564 (2459) &  685  (\sout{615}) & 195 (\sout{154}) \\
			                                        & 99\% & 26289 (26540) & 4312 (4247) & 1127 (\sout{1062}) & 309 (\sout{266}) \\
		\hline
			 & & \multicolumn{4}{c|}{$p=1\%$ or $p=99\%$} \\
		\hline
			\multirow{3}{*}{$1-\alpha$ (Confidence)}& 90\% & 27799 (27056) &  4699 (4329) & 1283 (\sout{1083}) & 377 (\sout{271}) \\
			                                        & 95\% & 39053 (38415) &  6506 (6147) & 1741 (\sout{1537}) & 497 (\sout{384}) \\
			                                        & 99\% & 66724 (66349) & 10947 (10616) & 2863 (\sout{2654}) & 789 (\sout{664}) \\
		\hline
	\end{tabular}
	\caption{Sample size for the determination of a proportion given the accuracy ($\delta=kp$) and confidence level ($1-\alpha$). The number without parenthesis
	   is the sample size calculated through the exact binomial formula, and in parenthesis the one calculated by the Gaussian approximation
		 (we have crossed out those sample
	   sizes for which the Gaussian approximation does not hold).}
	\label{tab:smallProportionSampleSize}
\end{table}	

\end{landscape}

\begin{landscape}

\begin{table}[htbp]
	\centering
	\begin{tabular}{|c|c|c|c|c|c|c|c|}
		\hline
			 & & \multicolumn{6}{c|}{$N$ (Sample size)} \\
		\hline
			 & & 5 & 10 & 15 & 20 & 25 & 30 \\
		\hline
			 & & \multicolumn{6}{c|}{ $\hat{p}=1\%$} \\
		\hline
			\multirow{3}{*}{$1-\alpha$ (Confidence)}& 90\% & [0,46.3] & [0,27.4] & [0,19.7] & [0,15.6] & [0,13.0] & [0,11.2] \\
			                                        & 95\% & [0,53.4] & [0,32.4] & [0,23.5] & [0,18.6] & [0,15.5] & [0,13.4] \\
			                                        & 99\% & [0,66.4] & [0,42.7] & [0,31.5] & [0,25.2] & [0,21.0] & [0,18.2] \\
		\hline
			 & & \multicolumn{6}{c|}{ $\hat{p}=2.5\%$} \\
		\hline
			\multirow{3}{*}{$1-\alpha$ (Confidence)}& 90\% & [0,48.1] & [0,29.6] & [0,22.1] & [0,18.0] & [0,15.4] & [0,13.6] \\
			                                        & 95\% & [0,55.1] & [0.34.6] & [0,25.9] & [0,21.1] & [0,18.0] & [0,15.9] \\
			                                        & 99\% & [0,67.8] & [0,44.9] & [0,34.0] & [0.27.8] & [0.23.7] & [0,20.9] \\
		\hline
			 & & \multicolumn{6}{c|}{ $\hat{p}=5\%$} \\
		\hline
			\multirow{3}{*}{$1-\alpha$ (Confidence)}& 90\% & [0,51.0] & [0,33.1] & [0.1,25.7] & [0.3,21.6] & [0.4,19.0] & [0.6,17.3]  \\
			                                        & 95\% & [0,57.8] & [0,38.1] & [0,29.6]   & [0.1,24.9] & [0.2,21.8] & [0.4,19.7] \\
			                                        & 99\% & [0,70.2] & [0,48.3] & [0,37.8]   & [0,31.7]   & [0.1,27.7] & [0.1,24.9] \\
		\hline
			 & & \multicolumn{6}{c|}{ $\hat{p}=95\%$} \\
		\hline
			\multirow{3}{*}{$1-\alpha$ (Confidence)}& 90\% & [49.0,100] & [66.9,100] & [74.3,99.9] & [78.4,99.7] & [81.0,99.6] & [82.7,99.4] \\
			                                        & 95\% & [42.2,100] & [61.9,100] & [70.4,100] & [75.1,99.9] & [78.2,99.8] & [80.3,99.6] \\
			                                        & 99\% & [29.8,100] & [51.7,100] & [62.2,100] & [68.3,100] & [72.3,99.9] & [75.1,99.9] \\
		\hline
			 & & \multicolumn{6}{c|}{ $\hat{p}=97.5\%$} \\
		\hline
			\multirow{3}{*}{$1-\alpha$ (Confidence)}& 90\% & [51.9,100] & [70.4,100] & [77.9,100] & [82.0,100] & [84.6,100] & [86.4,99.9] \\
			                                        & 95\% & [44.9,100] & [65.4,100] & [74.1,100] & [78.9,100] & [82.0,100] & [84.1,100] \\
			                                        & 99\% & [32.2,100] & [55.1,100] & [66.0,100] & [72.2,100] & [76.3,100] & [79.1,100] \\
		\hline
			 & & \multicolumn{6}{c|}{ $\hat{p}=99\%$} \\
		\hline
			\multirow{3}{*}{$1-\alpha$ (Confidence)}& 90\% & [53.7,100] & [72.6,100] & [80.3,100] & [84.4,100] & [87.0,100] & [88.8,100] \\
			                                        & 95\% & [46.6,100] & [67.6,100] & [76.5,100] & [81.4,100] & [84.5,100] & [86.6,100] \\
			                                        & 99\% & [33.6,100] & [57.3,100] & [68.5,100] & [74.8,100] & [79.0,100] & [81.8,100] \\
		\hline
	\end{tabular}
	\caption{Proportion confidence interval (in \%) given the sample size ($N$) and confidence level ($1-\alpha$). Because of the low $N$, the Gaussian approximation does not hold for most of the cells and the confidence intervals have been calculated using the exact Clopper-Pearson formula (note that the confidence interval does not need to be symmetric).}
	\label{tab:smallProportionAccuracy}
\end{table}	
\end{landscape}

\begin{landscape}

\begin{table}[htbp]
	\centering
	\begin{tabular}{|c|c|c|c|c|}
		\hline
			 & & Clopper-Pearson & Zero acceptance & $\chi^2$ approximation \\
		\hline
			 & & \multicolumn{3}{c}{$p_U=5\%$ or $p_L=95\%$} \\
		\hline
			\multirow{3}{*}{$1-\alpha$ (Confidence)}& 90\% & 45 & 45 & 47 \\
			                                        & 95\% & 59 & 59 & 60 \\
			                                        & 99\% & 90 & 90 & 93 \\
		\hline
			 & & \multicolumn{3}{c}{$p=2.5\%$ or $p=97.5\%$} \\
		\hline
			\multirow{3}{*}{$1-\alpha$ (Confidence)}& 90\% & 91 & 91 & 93 \\
			                                        & 95\% & 119 & 119 & 120 \\
			                                        & 99\% & 182 & 182 & 182\\
		\hline
			 & & \multicolumn{3}{c}{$p=1\%$ or $p=99\%$} \\
		\hline
			\multirow{3}{*}{$1-\alpha$ (Confidence)}& 90\% & 230 & 230 & 231 \\
			                                        & 95\% & 299 & 299 & 300 \\
			                                        & 99\% & 459 & 459 & 461 \\
		\hline
	\end{tabular}
	\caption{Sample size for the determination of a proportion given the limit ($p_U$ or $p_L$) and confidence level ($1-\alpha$). The first column
	   is the sample size calculated through the exact binomial formula, the second one is the one calculated by the zero acceptance, and the third column is the sample size
		 calculated through the $\chi^2$ approximation.}
	\label{tab:smallProportionOneSidedSampleSize}
\end{table}	

\end{landscape}

\begin{table}[htbp]
	\centering
	\begin{tabular}{|c|c|c|c|c|c|c|}
		\hline
			 & & \multicolumn{5}{c|}{Accuracy, $\delta$} \\
		\hline
			 & & 0.1 & 0.2 & 0.3 & 0.4 & 0.5\\
		\hline
			 & & \multicolumn{5}{c|}{$\rho=0.1$ or $\rho=-0.1$}\\
		\hline
			\multirow{3}{*}{$1-\alpha$ (Confidence)}& 90\% & 1062 & 267 & 120 & 68 & 44 \\
			                                        & 95\% & 1507 & 378 & 168 & 95 & 61 \\
			                                        & 99\% & 2600 & 649 & 288 & 162 & 103 \\
		\hline
			 & & \multicolumn{5}{c|}{$\rho=0.2$ or $\rho=-0.2$} \\
		\hline
			\multirow{3}{*}{$1-\alpha$ (Confidence)}& 90\% &  999 & 251 & 113 &  64 & 42 \\
			                                        & 95\% & 1417 & 355 & 159 &  90 & 58 \\
			                                        & 99\% & 2445 & 611 & 271 & 152 & 97 \\
		\hline
			 & & \multicolumn{5}{c|}{$\rho=0.3$ or $\rho=-0.3$} \\
		\hline
			\multirow{3}{*}{$1-\alpha$ (Confidence)}& 90\% &  898 & 226 & 102 &  58 & 38 \\
			                                        & 95\% & 1274 & 320 & 143 &  81 & 53 \\
			                                        & 99\% & 2198 & 550 & 244 & 138 & 88 \\
		\hline
			 & & \multicolumn{5}{c|}{$\rho=0.4$ or $\rho=-0.4$} \\
		\hline
			\multirow{3}{*}{$1-\alpha$ (Confidence)}& 90\% &  766 & 193 &  87 &  50 & 33 \\
			                                        & 95\% & 1086 & 273 & 123 &  70 & 46 \\
			                                        & 99\% & 1874 & 469 & 209 & 118 & 76 \\
		\hline
			 & & \multicolumn{5}{c|}{$\rho=0.5$ or $\rho=-0.5$} \\
		\hline
			\multirow{3}{*}{$1-\alpha$ (Confidence)}& 90\% &  612 & 155 &  71 &  41 & 27 \\
			                                        & 95\% &  867 & 219 &  99 &  57 & 37 \\
			                                        & 99\% & 1495 & 376 & 168 &  96 & 62 \\
		\hline
			 & & \multicolumn{5}{c|}{$\rho=0.6$ or $\rho=-0.6$} \\
		\hline
			\multirow{3}{*}{$1-\alpha$ (Confidence)}& 90\% &  447 & 114 &  53 &  31 & 21 \\
			                                        & 95\% &  633 & 161 &  74 &  43 & 29 \\
			                                        & 99\% & 1091 & 276 & 125 &  72 & 47 \\
		\hline
			 & & \multicolumn{5}{c|}{$\rho=0.7$ or $\rho=-0.7$} \\
		\hline
			\multirow{3}{*}{$1-\alpha$ (Confidence)}& 90\% &  286 &  75 &  36 &  22 & 15 \\
			                                        & 95\% &  404 & 105 &  49 &  30 & 20 \\
			                                        & 99\% &  696 & 178 &  82 &  48 & 33 \\
		\hline
			 & & \multicolumn{5}{c|}{$\rho=0.8$ or $\rho=-0.8$} \\
		\hline
			\multirow{3}{*}{$1-\alpha$ (Confidence)}& 90\% &  145 &  40 &  21 &  14 & 10 \\
			                                        & 95\% &  205 &  56 &  28 &  18 & 13 \\
			                                        & 99\% &  351 &  93 &  45 &  28 & 20 \\
		\hline
			 & & \multicolumn{5}{c|}{$\rho=0.9$ or $\rho=-0.9$} \\
		\hline
			\multirow{3}{*}{$1-\alpha$ (Confidence)}& 90\% &   45 &  15 &  10 &   8 &  7 \\
			                                        & 95\% &   62 &  20 &  12 &   9 &  8 \\
			                                        & 99\% &  105 &  33 &  19 &  14 & 11 \\
		\hline
	\end{tabular}
	\caption{Sample size for the determination of a correlation given the accuracy ($\delta$) and confidence level ($1-\alpha$).}
	\label{tab:correlationSampleSize}
\end{table}

\begin{table}[htbp]
	\centering
	\begin{small}
	\hspace*{-1cm}\begin{tabular}{|c|c|c|c|c|c|c|c|}
		\hline
			 & & \multicolumn{6}{c|}{$N$ (Sample size)} \\
		\hline
			 & & 5 & 10 & 15 & 20 & 25 & 30 \\
		\hline
			 & & \multicolumn{6}{c|}{ $r=0.1$} \\
		\hline
			\multirow{3}{*}{$1-\alpha$ (Confidence)}& 90\% & [-0.79,0.85] & [-0.48,0.62] & [-0.36,0.52] & [-0.29,0.46] & [-0.25,0.42] & [-0.21,0.39] \\
			                                        & 95\% & [-0.86,0.90] & [-0.57,0.69] & [-0.44,0.58] & [-0.36,0.52] & [-0.31,0.48] & [-0.27,0.44] \\
			                                        & 99\% & [-0.94,0.96] & [-0.70,0.79] & [-0.57,0.69] & [-0.48,0.62] & [-0.42,0.57] & [-0.38,0.53] \\
		\hline
			 & & \multicolumn{6}{c|}{ $r=0.2$} \\
		\hline
			\multirow{3}{*}{$1-\alpha$ (Confidence)}& 90\% & [-0.74,0.88] & [-0.40,0.68] & [-0.27,0.59] & [-0.19,0.54] & [-0.15,0.50] & [-0.11,0.48] \\
			                                        & 95\% & [-0.83,0.92] & [-0.49,0.74] & [-0.35,0.65] & [-0.27,0.59] & [-0.21,0.55] & [-0.17,0.52] \\
			                                        & 99\% & [-0.92,0.97] & [-0.65,0.83] & [-0.49,0.74] & [-0.40,0.68] & [-0.33,0.64] & [-0.29,0.60] \\
		\hline
			 & & \multicolumn{6}{c|}{ $r=0.3$} \\
		\hline
			\multirow{3}{*}{$1-\alpha$ (Confidence)}& 90\% & [-0.69,0.90] & [-0.30,0.73] & [-0.16,0.66] & [-0.09,0.61] & [-0.04,0.58] & [-0.01,0.56] \\
			                                        & 95\% & [-0.79,0.94] & [-0.41,0.78] & [-0.25,0.71] & [-0.16,0.66] & [-0.11,0.62] & [-0.07,0.60] \\
			                                        & 99\% & [-0.91,0.97] & [-0.58,0.86] & [-0.41,0.78] & [-0.31,0.73] & [-0.24,0.70] & [-0.18,0.67] \\
		\hline
			 & & \multicolumn{6}{c|}{ $r=0.4$} \\
		\hline
			\multirow{3}{*}{$1-\alpha$ (Confidence)}& 90\% & [-0.63,0.92] & [-0.20,0.78] & [-0.05,0.72] &  [0.03,0.68] &  [0.07,0.65] &  [0.11,0.63] \\
			                                        & 95\% & [-0.75,0.95] & [-0.31,0.82] & [-0.14,0.76] & [-0.05,0.72] &  [0.01,0.69] &  [0.05,0.67] \\
			                                        & 99\% & [-0.89,0.98] & [-0.50,0.89] & [-0.31,0.82] & [-0.20,0.78] & [-0.13,0.75] & [-0.07,0.73]\\
		\hline
			 & & \multicolumn{6}{c|}{ $r=0.5$} \\
		\hline
			\multirow{3}{*}{$1-\alpha$ (Confidence)}& 90\% & [-0.55,0.94] & [-0.07,0.83] &  [0.07,0.77] &  [0.15,0.74] &  [0.20,0.72] &  [0.23,0.70] \\
			                                        & 95\% & [-0.68,0.96] & [-0.19,0.86] & [-0.02,0.81] &  [0.07,0.77] &  [0.13,0.75] &  [0.17,0.73] \\
			                                        & 99\% & [-0.85,0.98] & [-0.40,0.91] & [-0.19,0.86] & [-0.08,0.83] &  [0.00,0.80] &  [0.05,0.78] \\
		\hline
			 & & \multicolumn{6}{c|}{ $r=0.6$} \\
		\hline
			\multirow{3}{*}{$1-\alpha$ (Confidence)}& 90\% & [-0.44,0.95] &  [0.07,0.87] &  [0.22,0.82] &  [0.29,0.80] &  [0.33,0.78] &  [0.36,0.77] \\
			                                        & 95\% & [-0.60,0.97] & [-0.05,0.89] &  [0.13,0.85] &  [0.21,0.82] &  [0.27,0.80] &  [0.31,0.79] \\
			                                        & 99\% & [-0.81,0.99] & [-0.27,0.93] & [-0.05,0.89] &  [0.07,0.87] &  [0.14,0.85] &  [0.20,0.83] \\
		\hline
			 & & \multicolumn{6}{c|}{ $r=0.7$} \\
		\hline
			\multirow{3}{*}{$1-\alpha$ (Confidence)}& 90\% & [-0.29,0.97] &  [0.24,0.90] &  [0.37,0.87] &  [0.44,0.85] &  [0.48,0.84] &  [0.50,0.83] \\
			                                        & 95\% & [-0.48,0.98] &  [0.13,0.92] &  [0.29,0.89] &  [0.37,0.87] &  [0.42,0.86] &  [0.45,0.85] \\
			                                        & 99\% & [-0.74,0.99] & [-0.11,0.95] &  [0.12,0.92] &  [0.24,0.90] &  [0.31,0.89] &  [0.36,0.88] \\
		\hline
			 & & \multicolumn{6}{c|}{ $r=0.8$} \\
		\hline
			\multirow{3}{*}{$1-\alpha$ (Confidence)}& 90\% & [-0.06,0.98] &  [0.44,0.94] &  [0.55,0.92] &  [0.60,0.91] &  [0.63,0.90] &  [0.65,0.89] \\
			                                        & 95\% & [-0.28,0.99] &  [0.34,0.95] &  [0.49,0.93] &  [0.55,0.92] &  [0.59,0.91] &  [0.62,0.90] \\
			                                        & 99\% & [-0.61,0.99] &  [0.12,0.97] &  [0.34,0.95] &  [0.44,0.94] &  [0.50,0.93] &  [0.54,0.92] \\
		\hline
			 & & \multicolumn{6}{c|}{ $r=0.9$} \\
		\hline
			\multirow{3}{*}{$1-\alpha$ (Confidence)}& 90\% &  [0.30,0.99] &  [0.69,0.97] &  [0.76,0.96] &  [0.79,0.95] &  [0.81,0.95] &  [0.82,0.95] \\
			                                        & 95\% &  [0.09,0.99] &  [0.62,0.98] &  [0.72,0.97] &  [0.76,0.96] &  [0.78,0.96] &  [0.80,0.95] \\
			                                        & 99\% & [-0.34,1.00] &  [0.46,0.99] &  [0.62,0.98] &  [0.69,0.97] &  [0.73,0.97] &  [0.75,0.96] \\
		\hline
	\end{tabular}
	\end{small}
	\caption{Correlation confidence interval given the sample size ($N$) and confidence level ($1-\alpha$). $r$ is the observed correlation. The table is written only for positive correlations.
	    Confidence intervals for negative correlations are just the opposite (for example, the 99\% confidence interval for an observed correlation of $r=-0.9$ 
			obtained from $N=5$ samples is $[-1.00,0.34]$).}
	\label{tab:correlationAccuracy}
\end{table}

\begin{landscape}

\begin{table}[htbp]
	\centering
	\begin{tabular}{|c|c|c|c|c|c|c|}
		\hline
			 & & \multicolumn{5}{c|}{Accuracy, $k$} \\
		\hline
			 & & 0.1 & 0.2 & 0.3 & 0.4 & 0.5\\
		\hline
			\multirow{3}{*}{$1-\alpha$ (Confidence)}& 90\% & 1086 & 274 & 124 &  71 & 47 \\
			                                        & 95\% & 1541 & 388 & 175 & 100 & 66 \\
			                                        & 99\% & 2660 & 670 & 301 & 172 & 112 \\
		\hline
	\end{tabular}
	\caption{Number of observed events, $E$, for the determination of a mean lifetime given the accuracy ($\delta=k\hat{\theta}$) and confidence level ($1-\alpha$).}
	\label{tab:expSampleSize}
\end{table}	

\begin{table}[htbp]
	\centering
	\begin{small}
	\begin{tabular}{|c|c|c|c|c|c|c|c|}
		\hline
			 & & \multicolumn{6}{c|}{$E$ (Number of events)} \\
		\hline
			 & & 5 & 10 & 15 & 20 & 25 & 30 \\
		\hline
			\multirow{3}{*}{$1-\alpha$ (Confidence)}& 90\% & [0.55,2.54] & [0.64,1.84] & [0.69,1.62] & [0.72,1.51] & [0.74,1.44] & [0.76,1.39] \\
			                                        & 95\% & [0.49,3.08] & [0.59,2.09] & [0.64,1.79] & [0.67,1.64] & [0.70,1.55] & [0.72,1.48] \\
			                                        & 99\% & [0.40,4.64] & [0.50,2.69] & [0.56,2.18] & [0.60,1.93] & [0.63,1.79] & [0.65,1.69] \\
		\hline
	\end{tabular}
	\end{small}
	\caption{Normalized confidence interval given the number of events ($E$) and confidence level ($1-\alpha$).}
	\label{tab:expAccuracy}
\end{table}	

\end{landscape}

\end{document}